\documentclass[runningheads]{llncs}

\usepackage{eccv}

\usepackage{eccvabbrv}

\usepackage{graphicx}
\usepackage{booktabs}
\usepackage{multirow}
\usepackage{rotating}
\usepackage{bbm}
\usepackage[table]{xcolor}
\definecolor{lightblue}{RGB}{100,180,255}
\definecolor{lightorange}{RGB}{220,120,40}
\definecolor{lightgreen}{RGB}{50,160,40}

\usepackage{algorithm}
\usepackage{algorithmic}

\usepackage[accsupp]{axessibility}  

\definecolor{citecolor}{HTML}{0071bc}
\usepackage[pagebackref=false,breaklinks=true,letterpaper=true,colorlinks,citecolor=citecolor,bookmarks=false]{hyperref}
\usepackage{url}            

\usepackage{orcidlink}

\begin{document}

	\title{SMART: When is it Actually Worth Expanding a Speculative Tree?} 
	


\author{Lifu Wang\inst{1} \and
Pan Zhou\inst{1} \thanks{Corresponding Author}
}

\authorrunning{L. Wang et al.}

\institute{Singapore Management University, Singapore \\
\email{\{lifuwang, panzhou\}@smu.edu.sg}\\
}

\maketitle

\begin{abstract}
Tree-based speculative decoding accelerates autoregressive generation by verifying a branching tree of draft tokens in a single target-model forward pass. However, existing methods prioritize maximizing token-level likelihood or the number of accepted tokens while ignoring a critical ``efficiency paradox'': the computational overhead of drafting and verifying big trees can grow super-linearly, particularly at scale. This often leads to negative wall-clock speedup when batch sizes increase or hardware saturation limits are reached. To address this, we propose \textbf{SMART}, a \textbf{S}ystem-aware \textbf{M}arginal \textbf{A}nalysis framework for \textbf{R}untime \textbf{T}ree construction. SMART reformulates tree expansion as a hardware-aware optimization problem that directly maximizes end-to-end speedup. By applying a principled marginal benefit--cost rule at inference time, SMART expands a node only when its marginal benefit--cost ratio exceeds the tree-level speedup. SMART is training-free and serves as a plug-and-play controller for existing frameworks like MSD and EAGLE. Extensive evaluations across three MLLMs (e.g., LLaVA, Qwen2-VL) and four LLMs (e.g., Llama-3.1, DeepSeek-R1) demonstrate that SMART consistently outperforms state-of-the-art baselines. It delivers an average additional speedup of \textbf{20.0\%} for MLLMs and \textbf{15.4\%} for LLMs across compute-bound batching regimes and diverse GPU architectures without performance loss.

\end{abstract}

\begin{figure}[t]
	\centering
	\begin{subfigure}[b]{0.48\textwidth}
		\centering
		\includegraphics[width=\textwidth]{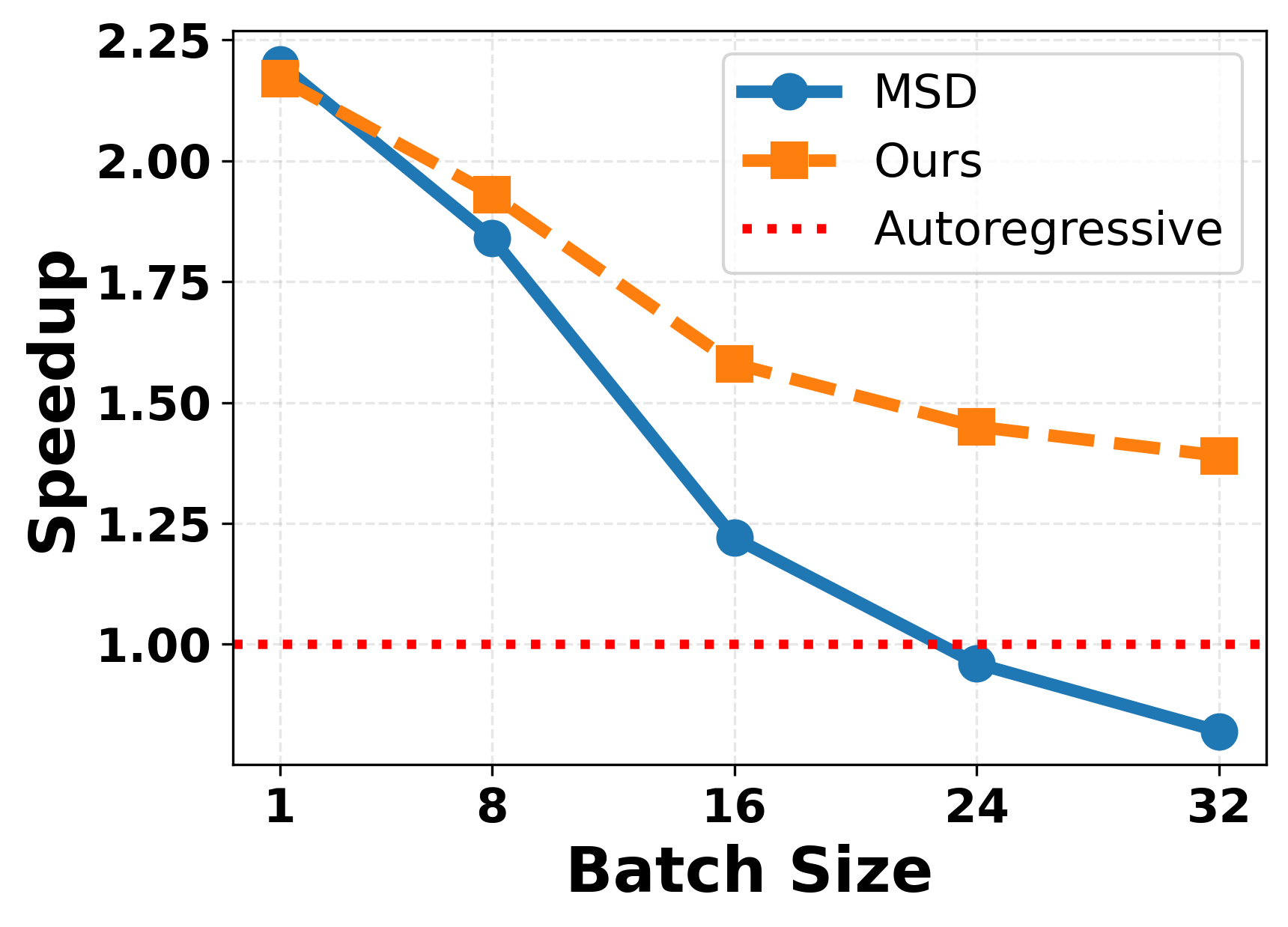}
		\caption{RTX Pro 6000}
		\label{fig:speedup-small}
	\end{subfigure}
	\hfill
	\begin{subfigure}[b]{0.48\textwidth}
		\centering
		\includegraphics[width=\textwidth]{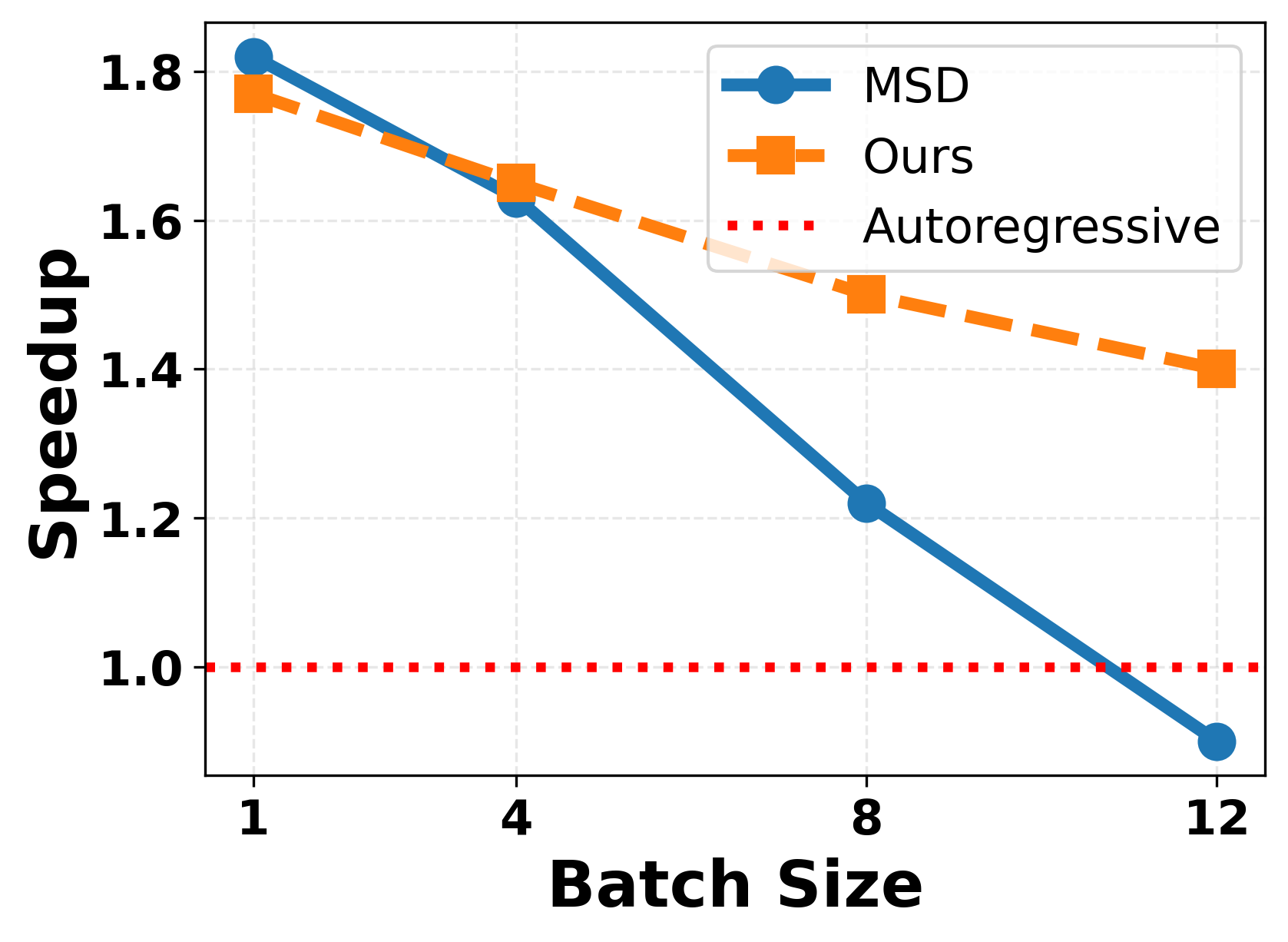}
		\caption{L40S}
		\label{fig:speedup-large}
	\end{subfigure}
	\caption{Speedup over autoregressive decoding across batch sizes (data from Table~\ref{tab:gpu-speedup}). Likelihood-maximizing tree methods such as Multimodal Speculative Decoding (MSD) exhibit severe performance degradation at large batch sizes, dropping below $1\times$ at batch 12 on L40S and reaching only $0.82\times$ at batch 32 on RTX Pro 6000. In contrast, our speedup-maximizing approach maintains consistent speedup by constructing trees based on the verification budget and device-specific cost models. This demonstrates that tree construction must be scalable with batch size and hardware-aware.}
	\label{fig:speedup-scaling}
\end{figure}

\section{Introduction}
\label{sec:intro}
Autoregressive decoding is the computational workhorse of modern generative AI, underpinning both Large Language Models (LLMs)~\cite{touvron2023llama, touvron2023llama2,grattafiori2024llama, achiam2023gpt, chiang2023vicuna} and Multimodal LLMs (MLLMs)~\cite{li2024llava,zhu2023minigpt, liu2023visual} for tasks ranging from (visual) reasoning~\cite{guo2025deepseek,chen2024spatialvlm} to high-fidelity image synthesis~\cite{sun2024autoregressive}. Yet its core mechanism is intrinsically sequential: tokens must be generated one after another. As model sizes grow and outputs become longer, this strict dependency chain turns into a dominant latency bottleneck, throttling throughput and inflating serving cost in real deployments.

Speculative decoding~\cite{leviathan2023fast,chen2023accelerating} has emerged as a practical strategy to break this sequential barrier. It uses a lightweight \emph{draft} model to propose multiple candidate tokens, which are then \emph{verified} by the target model in parallel. Tree-based speculative decoding~\cite{miao2024specinfer,li2025eagle,lin2025speculative} further extends this by constructing a \textbf{draft tree}---a branching structure of multiple candidate continuations. By verifying an entire tree in a single forward pass, the system increases the expected acceptance length (number of accepted tokens)  per target model forward. Conventional methods~\cite{li2025eagle,lin2025speculative} typically drive tree expansion via token-level likelihood, selecting candidates with the highest cumulative probability. 
More recently, GTO~\cite{hu2025bridging} argues that token-level likelihood maximization during training is a poor proxy for the actual speculative-decoding goal, and proposed training objectives that directly maximize the \emph{expected acceptance length} of the draft tree, aligning training with inference behavior and improving over vanilla speculative decoding.

 
Unfortunately,  the ultimate goal of speculative decoding is not to maximize the number of tokens accepted, but to maximize \textbf{end-to-end wall-clock speedup}. Existing designs suffer from a fundamental misalignment with system-level costs. Specifically, speedup is a function of both the acceptance length and the computational overhead of drafting and verification. Greedily expanding a tree to capture more tokens can be counterproductive; if the marginal cost of verifying a larger tree outweighs the gains in acceptance length, the system may experience ``negative speedup,'' performing worse than vanilla autoregressive decoding. Experienced inference practitioners are well aware of this issue and routinely profile latency to tune speculative-decoding configurations. However, current practice relies on offline grid search over static hyperparameters that remain constant across all inputs and must be re-tuned whenever the hardware, batch size, or workload distribution changes.

As illustrated in Fig.~\ref{fig:speedup-scaling}, this mismatch is exacerbated by two critical factors in production environments: \textbf{batch-size scalability} and \textbf{hardware heterogeneity}.  First, as batch sizes increase, the computational overhead of verifying big draft trees grows super-linearly. In memory-bandwidth bound regimes (e.g., $b=1$), verifying a large tree is beneficial as it amortizes the high cost of weight loading. However, once the batch size exceeds a hardware-specific threshold---approximately $b \ge 8$ for an RTX Pro 6000---the GPU shifts into a \textbf{compute-bound regime}.  In this state, the arithmetic intensity of verifying a large tree for every sequence in the batch exceeds the device's peak throughput, causing the verification cost to outweigh the gains in acceptance length. Second, the ``pivot point" where this bottleneck occurs is highly device-specific. For instance, a likelihood-maximizing tree constructed by MSD yielding $1.8\times$ speedup on an RTX Pro 6000 may drop to $1.2\times$ on an L40S at the same batch size of $8$ because the latter saturates its compute units earlier. These results underscore that a likelihood-maximizing tree is inherently suboptimal. Effective speculative decoding requires hardware-aware draft tree construction that adapts to the specific \textbf{arithmetic intensity} and \textbf{saturation limits} of the underlying GPU.

 In this paper, we adopt a \textbf{system-oriented} view of speculative decoding. Instead of asking \emph{how to maximize acceptance length}, we ask: 	\emph{When is it computationally worth expanding the draft tree, and which expansions measurably improve end-to-end speedup under the current hardware and batching regime?} Our goal is to automate the manual, latency-aware deployment tuning that practitioners currently perform offline into an \emph{online}, \emph{context-adaptive} marginal decision rule.

\noindent\textbf{Contributions.} We propose \textbf{SMART}, a \textbf{S}ystem-aware \textbf{M}arginal \textbf{A}nalysis framework for \textbf{R}untime \textbf{T}ree construction in speculative decoding. SMART constructs a speedup-maximizing draft tree \emph{at inference time} using a principled marginal benefit--cost rule. Importantly, SMART is \textbf{training-free}: it requires no changes to the draft model or target model weights, making it an \emph{out-of-the-box} drop-in improvement for existing speculative decoding pipelines (e.g. MSD~\cite{lin2025speculative} and EAGLE-3~\cite{li2025eagle}). Our main contributions are three-fold. 
 
First, we define a \textbf{system-level speedup objective.}   Given a draft tree $\mathcal{T}$, we explicitly model the end-to-end speedup as $
 	\mathcal{R}(\mathcal{T}) \;=\; \frac{c_T \cdot L^{\text{tree}}}{C_{\text{draft}} + C_{\text{verify}}}, $ 
 where $L^{\text{tree}}$ is the expected number of accepted tokens (i.e., acceptance length) of the tree $\mathcal{T}$, $c_T$ is the per-token cost of vanilla autoregressive decoding under the target model, and $C_{\text{draft}}$ and $C_{\text{verify}}$ are the total drafting and verification costs induced by the tree. Intuitively,  the $c_T \cdot L^{\text{tree}}$ denotes the cost of vanilla sequential decoding for generating $L^{\text{tree}}$ tokens, while $C_{\text{draft}} + C_{\text{verify}}$ is the cost of speculative decoding for getting $L^{\text{tree}}$ accepted tokens. Therefore, this formulation makes the central trade-off explicit: maximizing $L^{\text{tree}}$ alone can be suboptimal if it increases $C_{\text{draft}}+C_{\text{verify}}$ disproportionately.  By directly optimizing the ratio $\mathcal{R}(\mathcal{T})$, SMART targets the metric that matters in deployment: wall-clock speedup relative to vanilla decoding.

Second, SMART builds upon the reward $\mathcal{R}(\mathcal{T})$ to propose a speedup-maximizing tree expansion framework. To maximize the reward efficiently, SMART formulates tree construction as a sequence of speedup-maximizing expansion decisions. At each layer, we estimate the marginal gain and marginal cost of expanding each candidate node, and expand a node only when its marginal benefit--cost ratio exceeds the current tree's global ratio. This criterion ensures that local expansions improve the global speedup objective, allowing the tree shape to adapt to both the context difficulty and the available hardware budget.

Finally, SMART is training-free and applicable for plug-and-play deployment. 
SMART does not modify the draft model, the target model, or the verification mechanism. Instead, it replaces the likelihood-maximizing tree-construction policy with a speedup-maximizing policy. As a result, SMART is immediately compatible with existing speculative decoding systems and can be integrated as a lightweight inference-time controller.

Extensive evaluations across three MLLMs (LLaVA-1.5-7B and LLaVA-1.5-13B~\cite{liu2023visual}, and Qwen2-VL-7B-Instruct~\cite{Qwen2-VL}) and four LLMs (LLaMA-3.1-Instruct-8B and LLaMA-3.3-70B~\cite{grattafiori2024llama}, Vicuna-1.3-13B~\cite{chiang2023vicuna}, and DeepSeek-R1-Distill-LLaMA-8B~\cite{guo2025deepseek}) demonstrate that SMART consistently improves end-to-end speedup over strong baselines such as MSD~\cite{lin2025speculative} and EAGLE-3~\cite{li2025eagle}, yielding an average of \textbf{20.0\%} additional acceleration on MLLMs and \textbf{15.4\%} on LLMs, while remaining robust across diverse hardware and batching scenarios.

\section{Related Work}

Speculative decoding accelerates autoregressive generation by proposing draft tokens with a lightweight model and verifying them in parallel using the target model.~\cite{leviathan2023fast,chen2023accelerating,sun2023spectr,miao2024specinfer}. One line of work focuses on the design and training of draft models. Medusa~\cite{cai2024medusa} trains multiple prediction heads to generate draft tokens in parallel. EAGLE~\cite{li2024eagle} predicts future representations in the feature space, while EAGLE-3~\cite{li2025eagle} later returns to token-level prediction with scaled training data. HASS~\cite{zhang2024learning} explicitly mitigates feature-level draft–target mismatches, and GRIFFIN~\cite{hu2025griffin} further reduces token-level draft–target mismatches. GTO~\cite{hu2025bridging} treats the expected acceptance length of draft trees as a reward signal and updates the draft model using a PPO-style surrogate objective. Another line of work focuses on constructing draft trees, which is also the focus of our work. These methods can be divided based on whether they require training extra modules. On the training side, SpecDec++~\cite{huang2024specdec++} and DISCO~\cite{mamou2024dynamic} learn classifiers to predict optimal draft lengths. On the inference side, EAGLE-2~\cite{li2024eagle2} proposes a context-aware dynamic draft tree to increase acceptance length. TapOut~\cite{sridhar2025tapout} and SVIP~\cite{zhang2025draft} rely on per-token heuristics such as entropy or confidence scores to determine when to stop drafting. However, although these token-level heuristics are simple and system-friendly, they often suffer from threshold sensitivity and limited transferability. In contrast, SMART makes expansion decisions through a marginal benefit–cost ratio analysis and does not rely on externally tuned thresholds.

\begin{figure}[t]
\centering

\noindent\makebox[0.66\textwidth][c]{Likelihood-Maximizing Tree}\hfill
\makebox[0.32\textwidth][c]{Speedup-Maximizing Tree}\\

\begin{subfigure}[t]{0.32\textwidth}
    \centering
    \includegraphics[width=\textwidth]{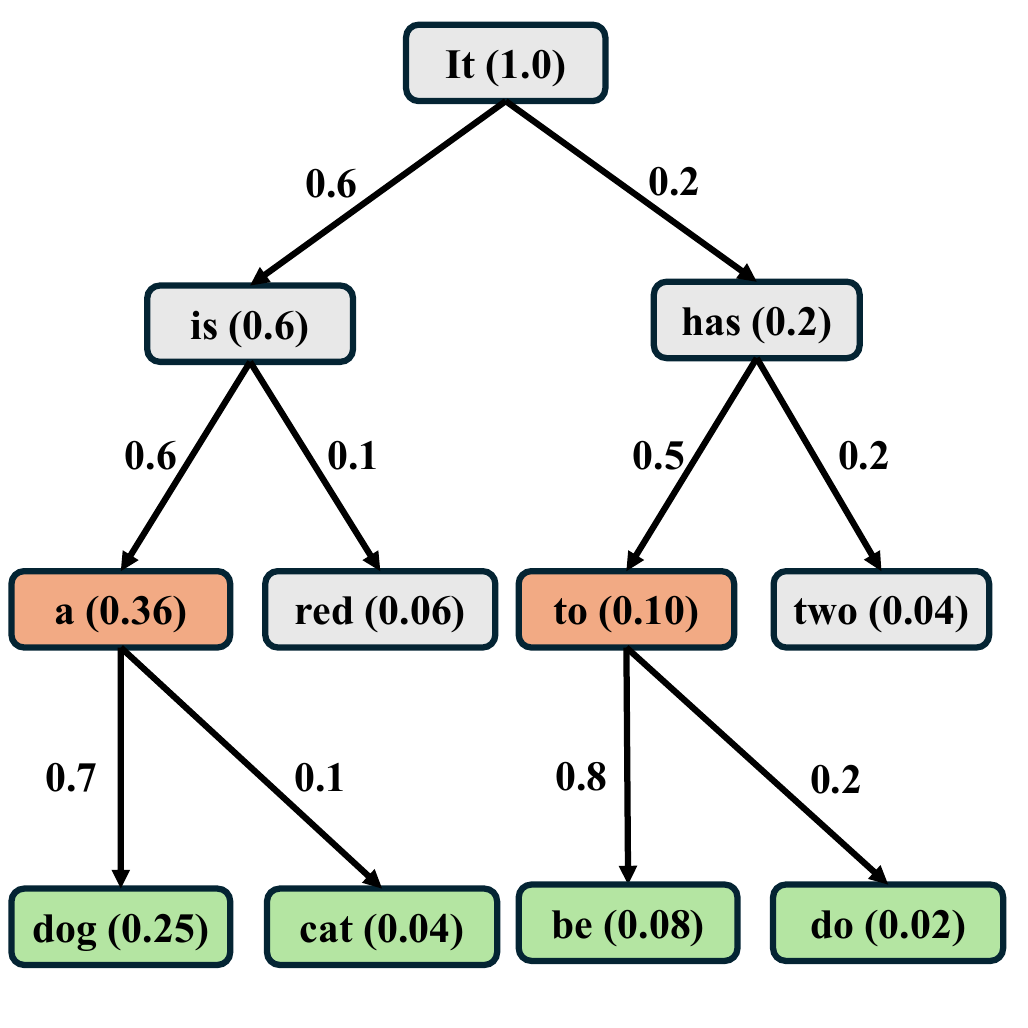}
    \caption{Expand with Top-2}
    \label{fig:method-expand}
\end{subfigure}
\hfill
\begin{subfigure}[t]{0.32\textwidth}
    \centering
    \includegraphics[width=\textwidth]{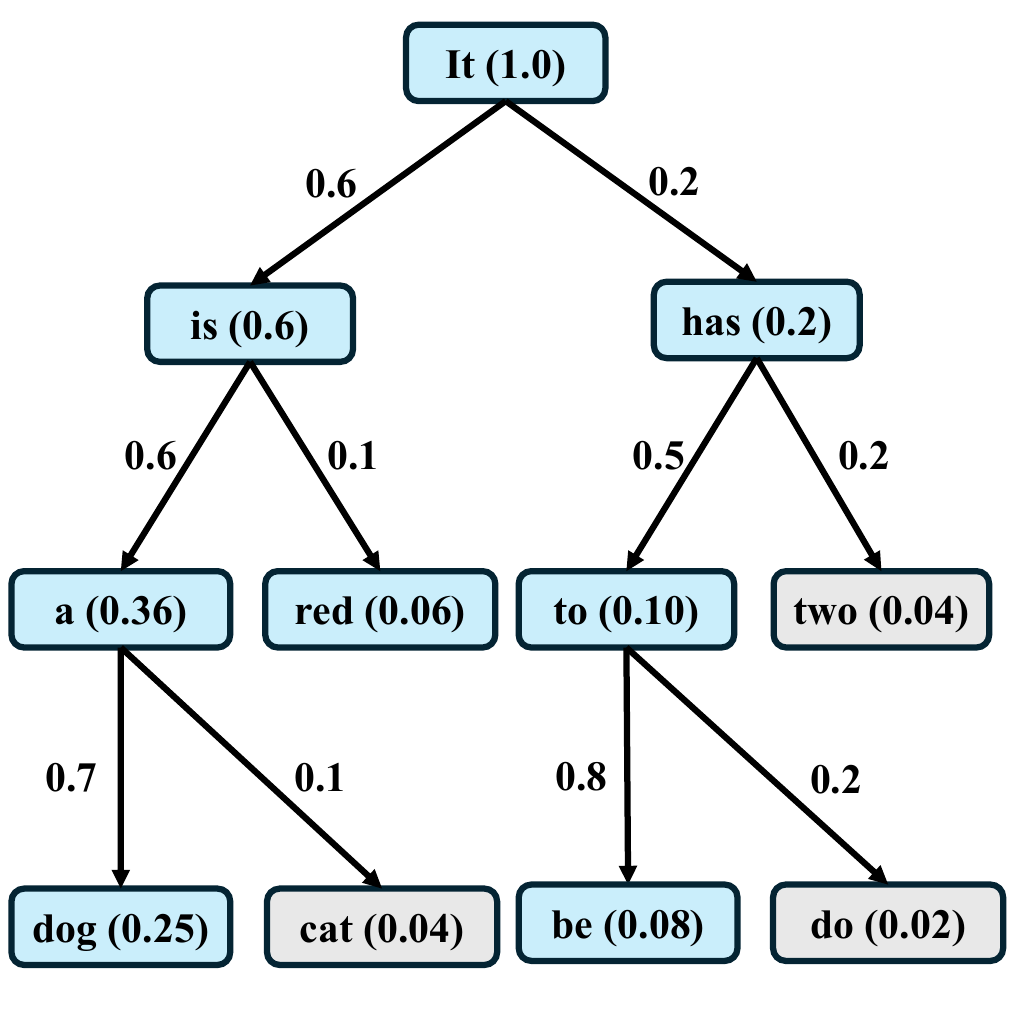}
    \caption{Rerank with Top-8}
    \label{fig:method-rerank}
\end{subfigure}
\hfill\vrule width 0.6pt\hfill 
\begin{subfigure}[t]{0.32\textwidth}
    \centering
    \includegraphics[width=\textwidth]{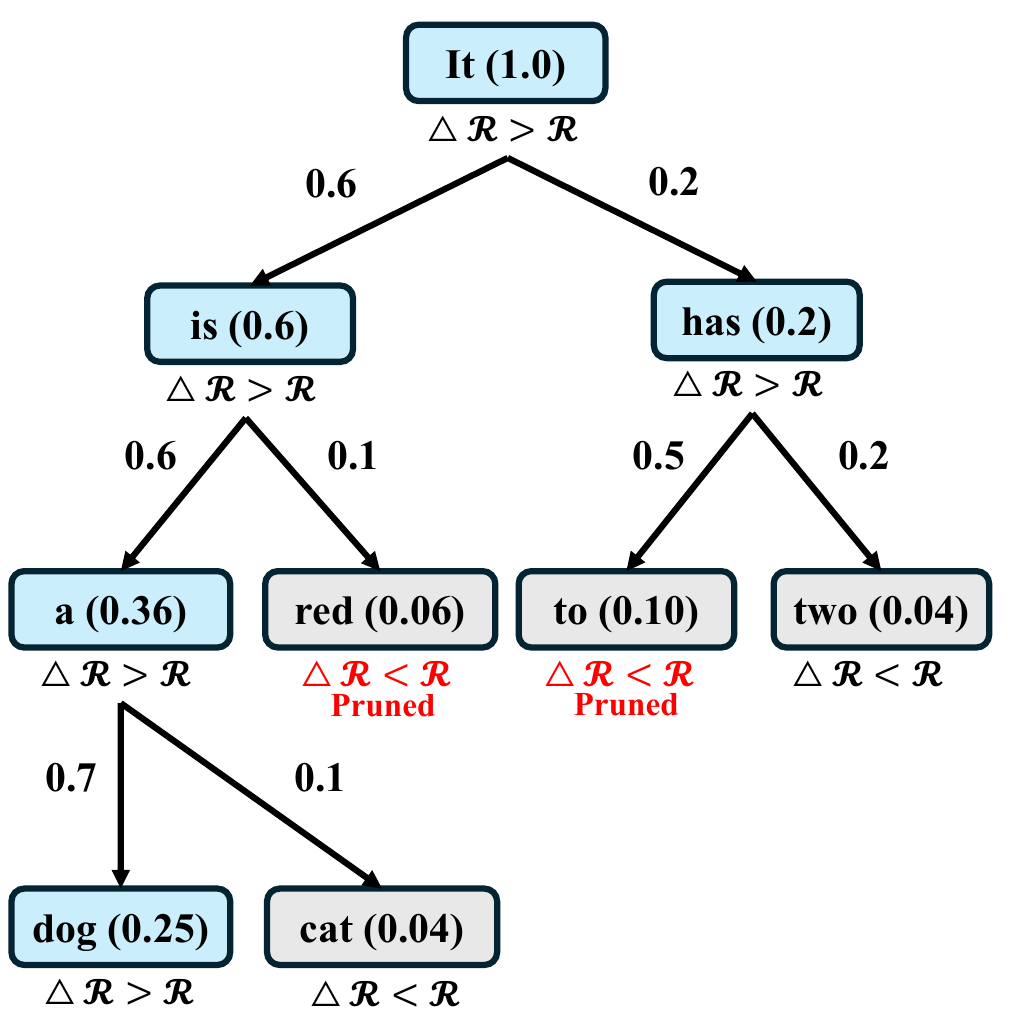}
    \caption{SMART}
    \label{fig:smart-tree}
\end{subfigure}

\caption{
Comparison of likelihood-maximizing ((a)–(b)) and speedup-maximizing tree construction (c).
\textbf{(a) Expansion phase.} At each layer, the method selects the top-2 nodes with the highest cumulative probability predicted by the draft model (\textcolor{lightorange}{orange}) and generating their top-2 children (\textcolor{lightgreen}{green}) using the draft model.
\textbf{(b) Rerank phase.} After reaching the maximum depth, all nodes in the tree are globally reranked by confidence and the top-8 nodes (\textcolor{lightblue}{blue}) are retained for verification.
\textbf{(c) SMART.} Instead of expanding all candidates, SMART evaluates each node's marginal benefit--cost ratio ($\Delta R = \frac{\Delta C_{\text{target}}(u)}{\Delta C_{\text{spec}}(u)}$, with marginal terms defined in Eqn.~\ref{eq:definition_delta}) online and only expands nodes (\textcolor{lightblue}{blue}) that improve the overall speedup, producing a smaller and more efficient draft tree.
}

\label{fig:method-comparison}
\end{figure}

\section{Methodology}

Motivated by the mismatch between likelihood-driven tree heuristics and \emph{end-to-end} speedup, we propose a speedup-maximizing framework that constructs draft trees by directly optimizing a device-specific speedup objective. Our method has two components. First, we define an end-to-end speedup metric (Sec.~\ref{sec:static-reward}) that captures true wall-clock efficiency by balancing expected accepted tokens against the measured costs of drafting and verification on a given device. Second, we formulate tree construction as a sequential decision problem (Sec.~\ref{sec:adaptive-tree}) and greedily expand tree nodes only when their marginal benefit--cost ratio exceeds the current tree's global ratio. This ensures each tree expansion improves the global speedup objective while preserving linear-time construction.

\subsection{Expected Speedup of A Draft Tree}
\label{sec:static-reward}
We start by introducing how likelihood-maximizing draft trees are constructed. State-of-the-art approaches such as EAGLE-3~\cite{li2025eagle} and MSD~\cite{lin2025speculative} construct a depth-$d$ draft tree with a fixed two-stage policy: (i) as shown in Fig.~\ref{fig:method-expand}, at each layer, we select the global top-$k$ nodes by  computing cumulative probability $P(u)$ from token probability given by the draft model; (ii) as shown in Fig.~\ref{fig:method-rerank}, after reaching the maximal depth $d$, namely, finishing tree construction, we re-rank tree nodes by $P(\cdot)$ and keep the top-$g$ nodes for verification. These methods optimize the acceptance rate of draft tokens. However, higher acceptance does not necessarily imply higher speedup. As discussed in Sec.~\ref{sec:intro}, this assumption can fail because speedup depends on \emph{both} the number of tokens accepted and the computation cost required to produce and verify them, which varies with batch size and hardware. 

To resolve this issue, we  evaluate the quality of a draft tree using a system-level speedup that directly compares the generation cost (i.e., wall-clock time) of vanilla autoregressive decoding and speculative decoding. Formally, given draft tree $\mathcal{T}$, we define  its  \textbf{expected speedup} as
\begin{equation}\label{reward}
\mathcal{R}(\mathcal{T})=
\frac{c_T \cdot L^{\text{tree}}}{C_{\text{draft}}(\mathcal{T}) + C_{\text{verify}}(\mathcal{T})} 
\end{equation}
where $c_T$ is the per-token autoregressive decoding cost (i.e., wall-clock time) of the target model, $L^{\text{tree}}$ is the expected number of tokens accepted from $\mathcal{T}$, and $C_{\text{draft}}(\mathcal{T})$ and $C_{\text{verify}}(\mathcal{T})$ are the measured costs to generate and verify the draft tree, respectively. 
To generate $L^{\text{tree}}$ tokens, target model needs to forward $L^{\text{tree}}$ times and thus needs the total cost  $c_T \cdot L^{\text{tree}}$ which corresponds to the  numerator  in Eqn.~\eqref{reward}. Meanwhile, to obtain an expected acceptance length of $L^{\text{tree}}$, speculative decoding must generate a draft tree and verify it with the target model. The total cost is the sum of drafting and verification costs, corresponding to the denominator $C_{\text{draft}}(\mathcal{T}) + C_{\text{verify}}(\mathcal{T})$ in Eqn.~\eqref{reward}. Unlike likelihood- or acceptance-length-only objectives, $\mathcal{R}(\mathcal{T})$ directly measures the speedup by comparing  the cost of vanilla autoregressive decoding of the target model and the cost of speculative decoding under the same acceptance length. This new metric is hardware-aware that adapts to the specific {arithmetic intensity} and {saturation limits} of the underlying GPU.

\noindent{\textbf{Estimation of Expected Accepted Tokens $L^{\text{tree}}$}.}  To evaluate Eqn.~\eqref{reward}, we first estimate the tree acceptance length $L^{\text{tree}}$ of the draft tree $\mathcal{T}$, which in turn yields an estimate of the target model's sequential decoding cost: $c_T \cdot L^{\text{tree}}$. To this end, we follow the spirit of GTO~\cite{hu2025bridging} and compute the expected acceptance length $L^{\text{tree}}$ as the mean across all paths in the tree $\mathcal{T}$. Specifically, given a context $\mathbf{x}$, the draft model generates draft tokens to construct a tree $\mathcal{T}$  which contains $|\mathcal{P}|$ draft sequences, and the corresponding acceptance length $L^{\text{tree}}$ can be estimated as 
\begin{equation}
    \label{eq:expected_acc_length}
L^{\text{tree}} \;=\; \frac{1}{|\mathcal{P}|}\sum\nolimits_{\tilde{\mathbf{x}}^{(i)}\in\mathcal{P}} L_i = \frac{1}{|\mathcal{P  }|} \sum\nolimits_{\tilde{\mathbf{x}}^{(i)}\in\mathcal{P}} \sum\nolimits_{j=1}^{|\tilde{\mathbf{x}}^{(i)}|} P\!\left(\tilde{\mathbf{x}}_{1:j}^{(i)}\mid \mathbf{x} \right),
\end{equation}
where $\mathcal{P}$ denotes the set of all root-to-leaf paths in the  tree $\mathcal{T}$ and $P\big(\tilde{\mathbf{x}}_{1:j}^{(i)}\mid \mathbf{x} \big)$ is the accumulated probability of sequence $\tilde{\mathbf{x}}_{1:j}^{(i)}$:
\begin{equation}
	\label{eq:cum_prob}
P\!\left(\tilde{\mathbf{x}}_{1:j}^{(i)}\mid \mathbf{x} \right)
	\;=\;
	\prod\nolimits_{k=1}^{j} p\!\left(\tilde{\mathbf{x}}_{k}^{(i)}\mid \mathbf{x},\tilde{\mathbf{x}}_{1:k-1}^{(i)}\right),
\end{equation}
where $p\big(\tilde{\mathbf{x}}_{k}^{(i)}\mid \mathbf{x},\tilde{\mathbf{x}}_{1:k-1}^{(i)}\big)$ denotes the probability of generating token $\tilde{\mathbf{x}}_{k}^{(i)}$ of target model given context $[\mathbf{x},\tilde{\mathbf{x}}_{1:k-1}^{(i)}]$. Here, $L_i \;=\; \sum_{j=1}^{|\tilde{\mathbf{x}}^{(i)}|} P\big(\tilde{\mathbf{x}}^{(i)}\mid \mathbf{x},\tilde{\mathbf{x}}_{1:j-1}^{(i)}\big)$ denotes the estimated acceptance length of the $i$-th draft path $\tilde{\mathbf{x}}^{(i)}$ in $\mathcal{T}$. This follows from the fact that the expected number of consecutively accepted tokens equals the sum of the probabilities that each prefix is accepted during verification~\cite{li2024eagle2}. We approximate the acceptance probability $p\big(\tilde{\mathbf{x}}_{k}^{(i)}\mid \mathbf{x},\tilde{\mathbf{x}}_{1:k-1}^{(i)}\big)$ using the draft model's predicted probability   because 1) draft model is trained to align with the target model; and 2) EAGLE-2~\cite{li2024eagle2} shows there is a strong positive correlation of prediction behaviors between target model and its corresponding  draft model.

 \begin{figure*}[t]
    \centering
    \begin{subfigure}[t]{0.49\textwidth}
        \centering
        \includegraphics[width=\linewidth]{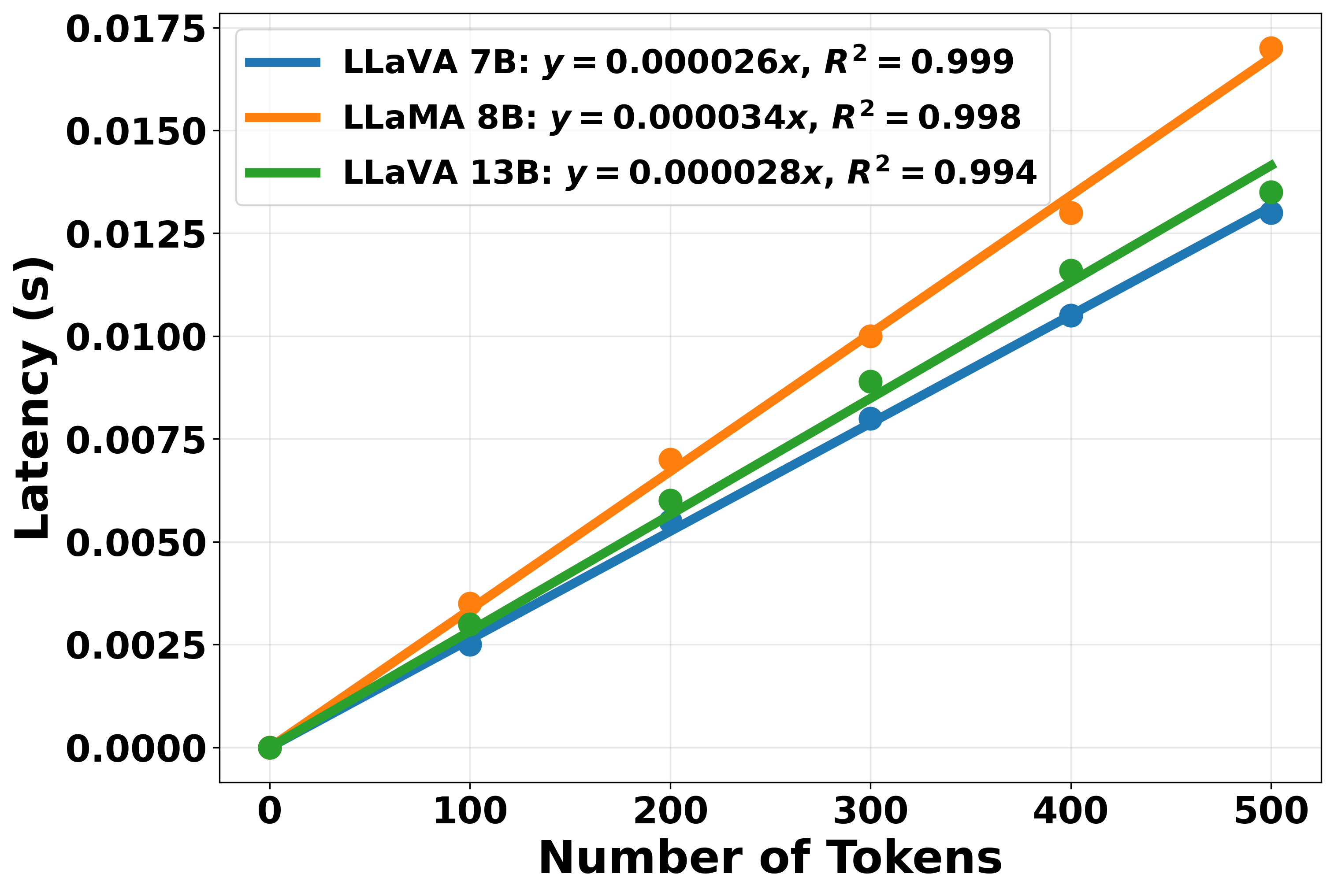}
        \caption{Draft cost vs.\ draft tokens $x$.}
        \label{fig:draft-cost}
    \end{subfigure}\hfill
    \begin{subfigure}[t]{0.49\textwidth}
        \centering
        \includegraphics[width=\linewidth]{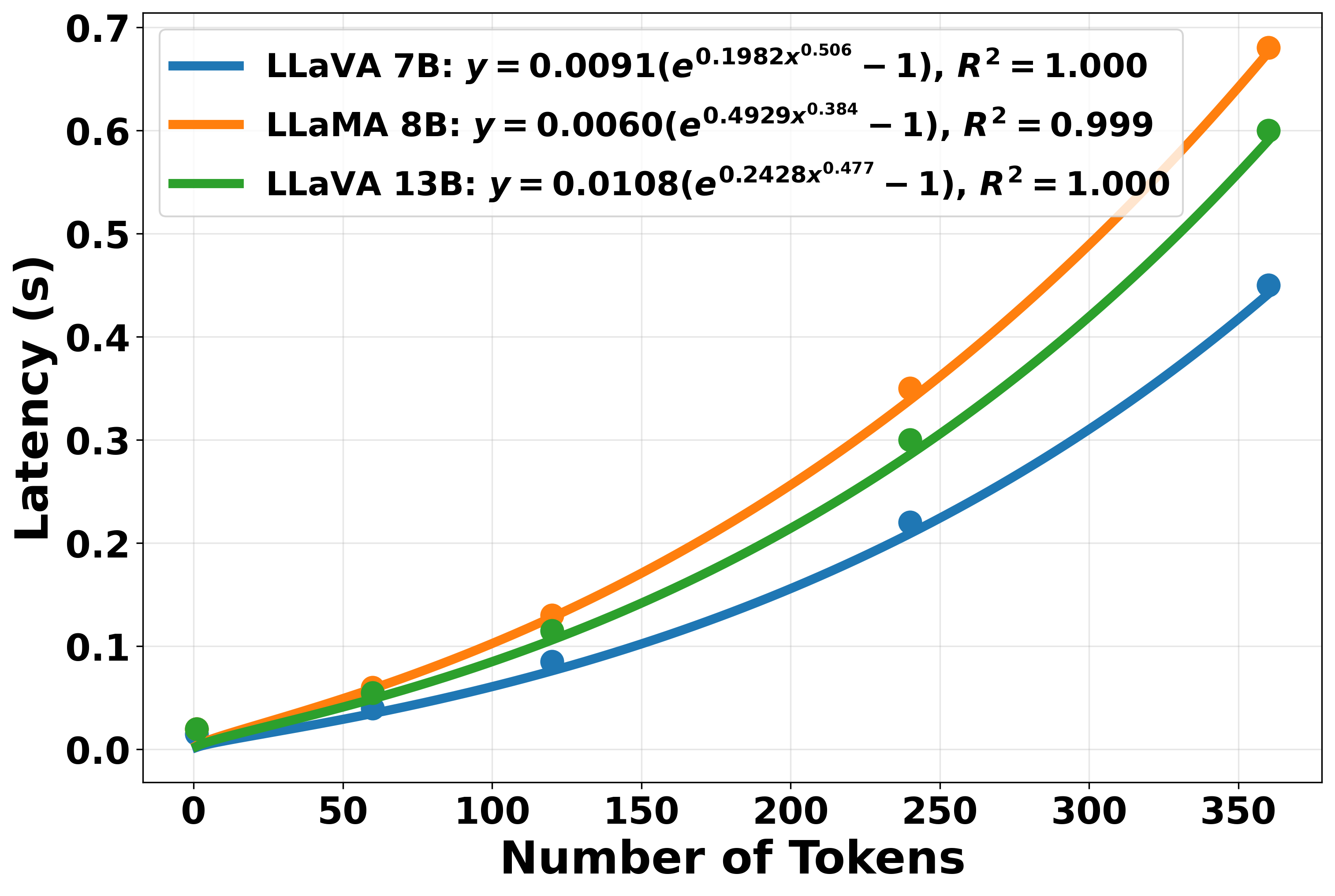}
        \caption{Verification cost vs.\ draft tokens $x$.}
        \label{fig:verify-cost}
    \end{subfigure}
    \caption{Measured latencies (dots) and fitted cost models (lines) for drafting and verification on RTX Pro 6000. Draft latency corresponds to the total latency of generating the full draft tree, where $x$ is the total number of tokens in the tree. Verification latency corresponds to the forward-pass latency of the target model with $x$ input tokens.}
    \label{fig:cost-models}
\end{figure*}

\subsubsection{Cost Modeling.}

As shown in Fig.~\ref{fig:cost-models}, we profile device-specific draft and verification latencies as a function of the total number of drafted tokens $|\mathcal{T}|$ in the tree $\mathcal{T}$. Drafting scales roughly linearly with $|\mathcal{T}|$ because draft models are small and typically memory-bound, yielding near-constant per-token cost, as demonstrated in Fig.~\ref{fig:cost-models}a. In practice, draft latency follows a layer-wise staircase: tokens at the same depth are generated in parallel, so cost increases only when a new depth level is added; the continuous form serves as a smooth surrogate for the marginal analysis. Verification grows much faster because the target model is large and, in high-batch compute-saturated regimes, verification becomes compute-bound, causing empirically convex latency growth as the number of draft tokens increases, as shown in Fig.~\ref{fig:cost-models}b.

Accordingly, we use linear model to model the draft cost in Fig.~\ref{fig:cost-models} (a):
\begin{equation}
\label{eq:linear_model}
C_{\text{draft}}( \mathcal{T}) = \lambda  |\mathcal{T}| + \beta,
\end{equation}
and adopt a power-exponential model to approximate the verification latency:
\begin{equation}
\label{eq:exp_model}
C_{\text{verify}}(\mathcal{T})
=
\gamma\!\left(\exp\!\left(\delta\,|\mathcal{T}|^{\rho}\right)-1\right)+\eta .
\end{equation}
Here $\lambda, \gamma, \delta,$ and $\rho$ are fitted per device, with the bias terms ($\beta$ and $\eta$) fixed to $0$ to ensure both models pass through the origin. This profiling is lightweight, making the estimation of the hyperparameters in Eqs.~\eqref{eq:linear_model} and \eqref{eq:exp_model} inexpensive. In practice, each fit requires only five forward passes. For example, profiling LLaMA-3.1-Instruct-8B~\cite{grattafiori2024llama} on MT-Bench~\cite{zheng2023judging} takes about 10 seconds on an RTX Pro 6000 GPU, which is negligible and accounts for only $\approx 1.67\%$ of the MT-Bench test set inference time. This enables fast fitting of the two cost models, $C_{\text{draft}}(\mathcal{T})$ and $C_{\text{verify}}(\mathcal{T})$, which in turn supports efficient tree expansion in the next section.


\subsection{Speedup-Maximizing Draft Tree}
\label{sec:adaptive-tree}




Draft-tree construction is a trade-off: expanding more nodes can increase the expected acceptance length, but it also raises drafting and verification cost, which may reduce end-to-end speedup. Therefore, we must carefully choose which tree nodes to expand so as to improve expected acceptance length while incurring minimal additional drafting and verification cost. We formulate tree growth as a sequential decision process that, at each layer, selects which candidates to keep by comparing their marginal benefit in expected acceptance length against their marginal system cost, thereby maximizing end-to-end speedup.

\subsubsection{Sequential Decision Formulation.}
As shown in Fig.~\ref{fig:smart-tree}, we construct the draft tree sequentially over
layers $\ell=1,\ldots,d$. Let $S_\ell$ denote the set of all selected nodes after
completing layer~$\ell$, with $S_0=\{\mathrm{root}\}$. We maintain a layer-wise active
set $A_\ell$ (with $A_0=\{\mathrm{root}\}$), whose elements are the nodes retained at
layer~$\ell$ and expanded at layer~$\ell{+}1$.

At layer~$\ell$, we expand every node in $A_{\ell-1}$ by drawing its top-$k$ candidates
from the draft-model distribution, producing the candidate set
$\mathcal{U}_\ell(A_{\ell-1})$ of size $k|A_{\ell-1}|$. For each candidate token
$u \in \mathcal{U}_\ell(A_{\ell-1})$, a selection operator $\mathcal{E}_\ell$ assigns a
binary label $e_\ell(u)\in\{0,1\}$, where $e_\ell(u)=1$ indicates that $u$ is retained
and $e_\ell(u)=0$ that it is pruned. The operator thus maps the full candidate set to the
subset of survivors:
\begin{equation}
    A_\ell
    \;=\; \mathcal{E}_\ell\!\bigl(\mathcal{U}_\ell(A_{\ell-1})\bigr)
    \;=\; \bigl\{\, u \in \mathcal{U}_\ell(A_{\ell-1}) \;\big|\; e_\ell(u) = 1 \,\bigr\}.
\end{equation}
The tree evolves as
\begin{equation}
    S_\ell \;=\; S_{\ell-1}\cup A_\ell.
\end{equation}

Let $L\le d$ be the terminal layer: either $L=d$, or $L$ is the first layer~$\ell$
such that $A_\ell=\emptyset$ or $|S_\ell|$ reaches a predetermined budget~$B$.
The induced tree is $\mathcal{T}=S_L$.
Our objective is to choose the selection operators $\{\mathcal{E}_\ell\}_{\ell=1}^{d}$
to maximize the final-tree reward:
\begin{equation}\label{eq:optimization}
\begin{aligned}
\max_{\{\mathcal{E}_\ell\}_{\ell=1}^{d}}\quad
&\frac{c_T\,L^{\mathrm{tree}}(S_L)}
     {C_{\mathrm{draft}}(S_L)+C_{\mathrm{verify}}(S_L)} \\[4pt]
\text{s.t.}\quad &|S_L| \;\le\; B,
\end{aligned}
\end{equation}
where $|S_L| \le B$ constrains the tree size. Because verification latency grows
superlinearly with the number of draft tokens (Fig.~\ref{fig:verify-cost}), we impose a
total verification budget $B_{\mathrm{verify}}$ to keep verification in the memory-bound
(near-flat) region of the cost curve, and split it evenly across the batch, giving a
per-sequence budget $B = B_{\mathrm{verify}}/b$ with $b$ the batch size.

Optimizing Eqn.~\eqref{eq:optimization} seeks selection operators that maximize the
expected speedup of speculative decoding. The objective is the ratio of the cost of
vanilla autoregressive decoding by the target model, $c_T\,L^{\mathrm{tree}}(S_L)$, to
the cost of speculative decoding, $C_{\mathrm{draft}}(S_L) + C_{\mathrm{verify}}(S_L)$,
where $S_L$ is the final draft tree. Since the tree is built layer by layer, this
optimization reduces to a sequence of layer-wise decisions
$\{\mathcal{E}_\ell\}_{\ell=1}^{d}$: at each layer, retain only the candidates that
maximize the tree's expected speedup. Next, we describe an efficient strategy to solve
this sequential decision problem.

\subsubsection{Optimizing the Sequential Decision Objective}
Computing the optimal action sequence in Eqn.~\eqref{eq:optimization} requires evaluating all valid subtrees of the full $k$-ary expansion tree. Since the total number of candidate nodes across $d$ layers is $\sum_{\ell=1}^d k^\ell = \mathcal{O}(k^d)$, the number of possible configurations grows as $\mathcal{O}(2^{k^d})$, making exhaustive search intractable. Instead, we propose a greedy policy that makes locally optimal decisions at each layer.

Specifically, we include a candidate node $u$ (i.e., set $e_\ell(u)=1$) only if it 
increases the reward, i.e., $\Delta\mathcal{R}(u) > 0$. The reward is defined as
\begin{equation}
\mathcal{R}(\mathcal{T}) \;=\; \frac{C_{\text{target}}}{C_{\text{spec}}},
\quad
C_{\text{target}} = c_T \cdot L^{\text{tree}}(\mathcal{T}),
\quad
C_{\text{spec}} = C_{\text{draft}}(\mathcal{T}) + C_{\text{verify}}(\mathcal{T}),
\label{eq:definition}
\end{equation}
and the marginal increments upon adding $u$ are
\begin{equation}
\Delta C_{\text{target}}(u) \;=\; c_T \cdot \Delta L^{\text{tree}}(u),
\qquad
\Delta C_{\text{spec}}(u) \;=\; \Delta C_{\text{draft}}(u) + \Delta C_{\text{verify}}(u).
\label{eq:definition_delta}
\end{equation}
Working with the log-reward $J = \log \mathcal{R}(\mathcal{T}) = \log C_{\text{target}} 
- \log C_{\text{spec}}$, the change upon adding $u$ is
\begin{equation}
\Delta J(u)
= \log\!\left(1+\frac{\Delta C_{\text{target}}(u)}{C_{\text{target}}}\right)
-\log\!\left(1+\frac{\Delta C_{\text{spec}}(u)}{C_{\text{spec}}}\right)
\approx \frac{\Delta C_{\text{target}}(u)}{C_{\text{target}}}-\frac{\Delta C_{\text{spec}}(u)}{C_{\text{spec}}},
\end{equation}
using $\log(1+x)\approx x$ for small $x$. We include $u$ if and only if 
$\Delta J(u) > 0$, i.e.,
\begin{equation}
\Delta J(u) = \alpha\cdot \frac{\Delta C_{\text{target}}(u)}{\Delta C_{\text{spec}}(u)}
-
\frac{C_{\text{target}}}{C_{\text{spec}}} > 0,
\qquad \alpha\in(0,1],
\label{eq:local-global}
\end{equation}
which recovers the standard condition when $\alpha=1$. Here, $\alpha\in(0,1]$ 
accounts for optimistic draft-based acceptance estimates under draft--target mismatch.
The global terms $C_{\text{target}}$ and $C_{\text{spec}}$ are computed on the current 
tree using Eqn.~\eqref{eq:definition}. Next, we will compute the marginal terms in 
Eqn.~\eqref{eq:definition_delta} in 2 steps. 

\noindent\textbf{Step 1: Marginal benefit $\Delta C_{\text{target}}(u)$.}
To estimate $\Delta L^{\text{tree}}(u)$ efficiently, recall that $L^{\text{tree}}$
averages the expected acceptance length over all root-to-leaf paths
(Eqn.~\eqref{eq:expected_acc_length}).
Accordingly, the marginal benefit of expanding $u$ is diluted by $|\mathcal{P}|$:
\begin{equation}
\Delta L^{\text{tree}}(u)
\;\approx\;
\frac{1}{|\mathcal{P}|}\,\Delta L(u),
\qquad
\Delta L(u) \;=\; P(\tilde{\mathbf{x}}_u \mid \mathrm{anc}(u)),
\label{eq:marginal_benefit}
\end{equation}
where $P(\tilde{\mathbf{x}}_u \mid \mathrm{anc}(u))$ is the cumulative acceptance 
probability defined in Eqn.~\eqref{eq:cum_prob}.

\noindent\textbf{Step 2: Marginal cost $\Delta C_{\text{spec}}(u)$.}
We approximate the cost of adding one node ($\Delta n=1$) by differentiating the 
fitted cost models with respect to the current tree size $|\mathcal{T}|$:
\begin{equation}
\Delta C_{\text{spec}}(u)\approx C'_{\text{draft}}(|\mathcal{T}|)+C'_{\text{verify}}(|\mathcal{T}|).
\end{equation}
With
$C_{\text{draft}}(\mathcal{T})=\lambda |\mathcal{T}|+\beta$
and
$C_{\text{verify}}(\mathcal{T})=\gamma\!\left(\exp\!\left(\delta |\mathcal{T}|^{\rho}\right)-1\right)+\eta$,
this yields
\begin{equation}
\Delta C_{\text{spec}}(u)
\approx
\lambda
+
\gamma \delta \rho |\mathcal{T}|^{\rho-1}
\exp\!\left(\delta |\mathcal{T}|^{\rho}\right).
\label{eq:marginal_cost}
\end{equation}

Since we have calculated the marginal and global terms, we can decide whether to keep 
(expand) a candidate node $u$ by checking $\Delta J(u) > 0$:
\begin{equation}
e_\ell(u)\;=\;
\begin{cases}
1, & \text{if }\;\Delta J(u)
    \;=\;\alpha\cdot\dfrac{\Delta C_{\text{target}}(u)}{\Delta C_{\text{spec}}(u)}
    -\dfrac{C_{\text{target}}}{C_{\text{spec}}} \;>\; 0,\\[6pt]
0, & \text{otherwise.}
\end{cases}
\label{eq:decision_rule}
\end{equation}

Now we present the full SMART algorithm. At each layer $\ell$, we (i) generate top-$k$ candidates per parent, (ii) compute each candidate’s marginal benefit (Eqn.~\eqref{eq:marginal_benefit}) and marginal cost (Eqn.~\eqref{eq:marginal_cost}), (iii) evaluate the current tree-level reward (Eqn.~\eqref{eq:definition}), and (iv) apply the decision rule (Eqn.~\eqref{eq:decision_rule}) to determine which nodes to expand. The process continues until reaching the maximum depth $d$, exhausting the budget $B$, or when no candidate remains (i.e., $A_\ell=\emptyset$).

The greedy policy runs in $\mathcal{O}(kB)$ time: at each layer $\ell$, we evaluate 
$\Delta J(u)$ in $\mathcal{O}(1)$ for each of the $k|A_{\ell-1}|$ candidates, and 
the total number of candidates across all layers is bounded by $kB$. This reduces the 
complexity from $\mathcal{O}(2^{k^{d}})$ for exhaustive search to linear in the 
verification budget $B$. While not globally optimal, the greedy policy produces 
high-quality trees by ensuring each expansion locally improves the reward. 
Algorithm~1 in the appendix summarizes our greedy construction procedure.

	\section{Experiments}
	\label{sec:exp}

	\begin{table*}[t]
		\centering
		\caption{Comparison of speedup ratio $SR$ and acceptance rate $\beta$ on standard MLLM benchmarks with temperature $T \in \{0, 1\}$. The subscripts denote the relative improvement compared to the corresponding baseline. For example, at $T=0$, MSD+SMART on LLaVA-1.5 7B achieves the average $SR$ of 1.53 with an additional +29.7\% gain over the MSD value of 1.18.}
		\label{tab:mllm_speedup}
		\resizebox{\textwidth}{!}{%
			\begin{tabular}{c c c || cc cc cc cc cc cc | ll}
				\toprule
				& & & \multicolumn{2}{c}{VQAv2} & \multicolumn{2}{c}{AI2D} & \multicolumn{2}{c}{SQA Image} & \multicolumn{2}{c}{ChartQA} & \multicolumn{2}{c}{TextVQA} & \multicolumn{2}{c}{Hallusion} & \multicolumn{2}{c}{Average} \\
				\cmidrule(lr){4-5}\cmidrule(lr){6-7}\cmidrule(lr){8-9}\cmidrule(lr){10-11}\cmidrule(lr){12-13}\cmidrule(lr){14-15}\cmidrule(lr){16-17}
				& \textbf{Model} & \textbf{Method} & $SR$$\uparrow$ & $\beta$$\uparrow$ & $SR$$\uparrow$ & $\beta$$\uparrow$ & $SR$$\uparrow$ & $\beta$$\uparrow$ & $SR$$\uparrow$ & $\beta$$\uparrow$ & $SR$$\uparrow$ & $\beta$$\uparrow$ & $SR$$\uparrow$ & $\beta$$\uparrow$ & $SR$$\uparrow$ & $\beta$$\uparrow$ \\
				\midrule
				
				\multirow{15}{*}{\rotatebox[origin=c]{90}{Temperature = 0}}
				& \multirow{5}{*}{\shortstack{LLaVA-1.5\\7B}}
				& Medusa    & 0.88 & 0.48 & 0.82 & 0.44 & 0.86 & 0.46 & 0.91 & 0.51 & 0.95 & 0.54 & 1.01 & 0.58 & 0.91 & 0.50 \\
				& & EAGLE-1   & 0.95 & 0.52 & 0.88 & 0.48 & 0.92 & 0.50 & 0.98 & 0.55 & 1.02 & 0.58 & 1.08 & 0.62 & 0.97 & 0.54 \\
				& & EAGLE-2   & 1.08 & 0.59 & 0.98 & 0.53 & 1.02 & 0.54 & 1.06 & 0.62 & 1.15 & 0.64 & 1.18 & 0.68 & 1.08 & 0.60 \\
				\cmidrule(lr){3-17}
				& & MSD       & 1.23 & 0.67 & 1.09 & 0.58 & 1.09 & 0.56 & 1.14 & 0.68 & 1.26 & 0.68 & 1.26 & 0.71 & 1.18 & 0.65 \\
				& & \textbf{MSD+SMART} & \textbf{1.55} & \textbf{0.81} & \textbf{1.45} & \textbf{0.74} & \textbf{1.44} & \textbf{0.72} & \textbf{1.59} & \textbf{0.82} & \textbf{1.55} & \textbf{0.79} & \textbf{1.62} & \textbf{0.83} & $\mathbf{1.53_{+29.7\%}}$ & $\mathbf{0.79_{+21.5\%}}$ \\
				\cmidrule(lr){2-17}
				& \multirow{5}{*}{\shortstack{LLaVA-1.5\\13B}}
				& Medusa    & 0.95 & 0.44 & 0.88 & 0.41 & 1.01 & 0.46 & 1.10 & 0.54 & 0.91 & 0.42 & 0.88 & 0.44 & 0.96 & 0.45 \\
				& & EAGLE-1   & 1.02 & 0.48 & 0.95 & 0.45 & 1.08 & 0.50 & 1.18 & 0.58 & 0.98 & 0.46 & 0.95 & 0.48 & 1.03 & 0.49 \\
				& & EAGLE-2   & 1.15 & 0.54 & 1.08 & 0.50 & 1.18 & 0.54 & 1.32 & 0.62 & 1.12 & 0.51 & 1.08 & 0.51 & 1.16 & 0.54 \\
				\cmidrule(lr){3-17}
				& & MSD       & 1.30 & 0.59 & 1.17 & 0.53 & 1.28 & 0.57 & 1.45 & 0.66 & 1.22 & 0.54 & 1.17 & 0.54 & 1.26 & 0.57 \\
				& & \textbf{MSD+SMART} & \textbf{1.56} & \textbf{0.76} & \textbf{1.42} & \textbf{0.71} & \textbf{1.53} & \textbf{0.76} & \textbf{1.72} & \textbf{0.84} & \textbf{1.51} & \textbf{0.73} & \textbf{1.42} & \textbf{0.72} & $\mathbf{1.53_{+21.4\%}}$ & $\mathbf{0.75_{+31.6\%}}$ \\
				\cmidrule(lr){2-17}
				& \multirow{5}{*}{\shortstack{Qwen2VL\\7B Instruct}}
				& Medusa    & 0.85 & 0.54 & 0.79 & 0.48 & 0.82 & 0.47 & 0.98 & 0.54 & 0.88 & 0.50 & 0.91 & 0.54 & 0.87 & 0.51 \\
				& & EAGLE-1   & 0.92 & 0.58 & 0.85 & 0.52 & 0.88 & 0.51 & 1.05 & 0.58 & 0.95 & 0.54 & 0.98 & 0.58 & 0.94 & 0.55 \\
				& & EAGLE-2   & 1.05 & 0.65 & 0.96 & 0.56 & 0.98 & 0.55 & 1.15 & 0.62 & 1.08 & 0.58 & 1.08 & 0.61 & 1.05 & 0.60 \\
				\cmidrule(lr){3-17}
				& & MSD       & 1.18 & 0.71 & 1.05 & 0.58 & 1.06 & 0.57 & 1.24 & 0.65 & 1.16 & 0.60 & 1.15 & 0.63 & 1.14 & 0.62 \\
				& & \textbf{MSD+SMART} & \textbf{1.22} & \textbf{0.86} & \textbf{1.24} & \textbf{0.77} & \textbf{1.26} & \textbf{0.74} & \textbf{1.34} & \textbf{0.84} & \textbf{1.24} & \textbf{0.57} & \textbf{1.22} & \textbf{0.82} & $\mathbf{1.25_{+9.6\%}}$ & $\mathbf{0.77_{+24.2\%}}$ \\
				
				\midrule
				
				\multirow{15}{*}{\rotatebox[origin=c]{90}{Temperature = 1}}
				& \multirow{5}{*}{\shortstack{LLaVA-1.5\\7B}}
				& Medusa    & 1.74 & 0.24 & 1.43 & 0.28 & 1.58 & 0.29 & 1.33 & 0.26 & 0.97 & 0.19 & 1.46 & 0.26 & 1.42 & 0.25 \\
				& & EAGLE-1   & 1.85 & 0.26 & 1.52 & 0.30 & 1.68 & 0.31 & 1.42 & 0.28 & 1.05 & 0.21 & 1.55 & 0.28 & 1.51 & 0.27 \\
				& & EAGLE-2   & 2.05 & 0.28 & 1.65 & 0.32 & 1.85 & 0.33 & 1.52 & 0.29 & 1.15 & 0.23 & 1.68 & 0.30 & 1.65 & 0.29 \\
				\cmidrule(lr){3-17}
				& & MSD       & 2.21 & 0.30 & 1.77 & 0.34 & 2.00 & 0.35 & 1.63 & 0.31 & 1.22 & 0.24 & 1.77 & 0.31 & 1.77 & 0.31 \\
				& & \textbf{MSD+SMART} & \textbf{2.84} & \textbf{0.40} & \textbf{2.41} & \textbf{0.45} & \textbf{2.56} & \textbf{0.46} & \textbf{2.24} & \textbf{0.43} & \textbf{1.46} & \textbf{0.33} & \textbf{2.19} & \textbf{0.41} & $\mathbf{2.28_{+28.8\%}}$ & $\mathbf{0.42_{+35.5\%}}$ \\
				\cmidrule(lr){2-17}
				& \multirow{5}{*}{\shortstack{LLaVA-1.5\\13B}}
				& Medusa    & 1.67 & 0.27 & 1.39 & 0.30 & 1.91 & 0.33 & 1.36 & 0.26 & 1.07 & 0.20 & 1.23 & 0.24 & 1.44 & 0.27 \\
				& & EAGLE-1   & 1.78 & 0.29 & 1.48 & 0.32 & 2.02 & 0.35 & 1.45 & 0.28 & 1.15 & 0.22 & 1.32 & 0.26 & 1.53 & 0.29 \\
				& & EAGLE-2   & 1.95 & 0.31 & 1.58 & 0.33 & 2.25 & 0.36 & 1.58 & 0.29 & 1.25 & 0.24 & 1.42 & 0.27 & 1.67 & 0.30 \\
				\cmidrule(lr){3-17}
				& & MSD       & 2.14 & 0.33 & 1.71 & 0.35 & 2.40 & 0.38 & 1.71 & 0.31 & 1.33 & 0.25 & 1.48 & 0.28 & 1.79 & 0.31 \\
				& & \textbf{MSD+SMART} & \textbf{2.54} & \textbf{0.42} & \textbf{2.23} & \textbf{0.47} & \textbf{2.92} & \textbf{0.46} & \textbf{1.98} & \textbf{0.40} & \textbf{1.63} & \textbf{0.34} & \textbf{1.78} & \textbf{0.40} & $\mathbf{2.18_{+21.8\%}}$ & $\mathbf{0.41_{+32.3\%}}$ \\
				\cmidrule(lr){2-17}
				& \multirow{5}{*}{\shortstack{Qwen2VL\\7B Instruct}}
				& Medusa    & 1.24 & 0.52 & 0.91 & 0.38 & 1.49 & 0.42 & 0.95 & 0.42 & 1.14 & 0.42 & 0.88 & 0.42 & 1.10 & 0.43 \\
				& & EAGLE-1   & 1.32 & 0.55 & 0.98 & 0.41 & 1.58 & 0.45 & 1.02 & 0.45 & 1.22 & 0.45 & 0.95 & 0.45 & 1.18 & 0.46 \\
				& & EAGLE-2   & 1.42 & 0.59 & 1.06 & 0.43 & 1.68 & 0.47 & 1.08 & 0.46 & 1.32 & 0.47 & 1.02 & 0.47 & 1.26 & 0.48 \\
				\cmidrule(lr){3-17}
				& & MSD       & 1.50 & 0.62 & 1.12 & 0.44 & 1.79 & 0.48 & 1.15 & 0.48 & 1.38 & 0.48 & 1.07 & 0.48 & 1.33 & 0.50 \\
				& & \textbf{MSD+SMART} & \textbf{1.69} & \textbf{0.65} & \textbf{1.25} & \textbf{0.53} & \textbf{2.04} & \textbf{0.55} & \textbf{1.16} & \textbf{0.58} & \textbf{1.45} & \textbf{0.56} & \textbf{1.10} & \textbf{0.50} & $\mathbf{1.45_{+9.0\%}}$ & $\mathbf{0.56_{+12.0\%}}$ \\
				
				\bottomrule
			\end{tabular}%
		}
	\end{table*}
	

\noindent\textbf{Models \& Datasets.}
We evaluate SMART across a diverse range of models to demonstrate its generalizability. For \textbf{MLLMs}, we evaluate LLaVA-1.5 (7B/13B)~\cite{liu2023visual} and Qwen2VL-7B-Instruct~\cite{Qwen2-VL} on widely used multimodal benchmarks: VQAv2~\cite{antol2015vqa}, AI2D~\cite{kembhavi2016diagram}, ScienceQA\cite{lu2022learn}, ChartQA\cite{masry2022chartqa}, TextVQA~\cite{singh2019towards}, and HallusionBench\cite{guan2024hallusionbench}. For  \textbf{LLMs}, we report results on LLaMA-3.1-Instruct-8B~\cite{grattafiori2024llama}, Vicuna-1.3-13B~\cite{chiang2023vicuna}, DeepSeek-R1-Distill-LLaMA-8B~\cite{guo2025deepseek}, and LLaMA-3.3-70B~\cite{grattafiori2024llama} across three standard benchmarks covering chat, coding, and reasoning: MT-Bench~\cite{zheng2023judging}, HumanEval~\cite{chen2021evaluating}, and GSM8K~\cite{cobbe2021training}.

\noindent\textbf{Baselines.} 
For \textbf{MLLMs}, we integrate SMART into frameworks that utilize likelihood-maximizing tree construction (e.g., MSD~\cite{lin2025speculative}) and compare against Medusa~\cite{cai2024medusa}, EAGLE~\cite{li2024eagle}, EAGLE-2~\cite{li2024eagle2}, and MSD. For \textbf{LLMs}, we integrate SMART into Eagle-3~\cite{li2025eagle} and compare against a broad suite of representative baselines, including SPS~\cite{leviathan2023fast}, EAGLE, EAGLE-2, GRIFFIN~\cite{hu2025griffin}, and EAGLE-3. For a fair comparison, all methods are evaluated under the same hardware (RTX Pro 6000 GPUs) and decoding configurations. Due to limited space, we report results for integrating SMART into additional baselines in the appendix.

\noindent\textbf{Metrics \& Evaluation Protocol.}
Following prior work, we evaluate performance at decoding temperatures $T \in \{0, 1\}$. Since SMART is mathematically lossless, our evaluation focuses on efficiency via two key metrics. \textbf{(1) Speedup Ratio ($SR$):} The end-to-end wall-clock latency improvement relative to vanilla autoregressive decoding ($SR=1.00\times$). \textbf{(2) Acceptance Rate ($\beta$):} The fraction of drafted tokens accepted during verification.  We specifically report the acceptance \textit{rate} rather than the acceptance \textit{length}. Since SMART's pruning induces variable draft lengths across steps, raw per-step accepted token counts are no longer directly comparable; a normalized rate provides a more consistent measure of the draft model's efficiency relative to the chosen tree size.

	\begin{table*}[t]
		\centering
		\caption{
            Comparison of speedup ratio ($SR$) and acceptance rate ($\beta$) on standard LLM benchmarks under temperature settings $T \in \{0,1\}$. Subscripts denote the relative improvement over the corresponding baseline (EAGLE-3). For example, at $T=0$ on LLaMA-3.1-Instruct-8B, EAGLE-3+SMART achieves an average $SR$ of $1.59$, representing a +16.9\% improvement over the EAGLE-3 baseline value of $1.36$. 
            }
		\label{tab:speedup}
		\resizebox{\textwidth}{!}{%
			\begin{tabular}{cc||cccccc|ll||cccccc|ll}
				\toprule
				& & \multicolumn{8}{c||}{\textbf{Temperature = 0}} & \multicolumn{8}{c}{\textbf{Temperature = 1}} \\
				\noalign{\smallskip}
				\cmidrule(lr){3-10} \cmidrule(lr){11-18}
				\textbf{Model} & \textbf{Method} & \multicolumn{2}{c}{MT-bench} & \multicolumn{2}{c}{HumanEval} & \multicolumn{2}{c|}{GSM8K} & \multicolumn{2}{c||}{Average} & \multicolumn{2}{c}{MT-bench} & \multicolumn{2}{c}{HumanEval} & \multicolumn{2}{c|}{GSM8K} & \multicolumn{2}{c}{Average} \\
				\cmidrule(lr){3-10} \cmidrule(lr){11-18}
				& & $SR$$\uparrow$ & $\beta$$\uparrow$ & $SR$$\uparrow$ & $\beta$$\uparrow$ & $SR$$\uparrow$ & $\beta$$\uparrow$ & $SR$$\uparrow$ & $\beta$$\uparrow$ & $SR$$\uparrow$ & $\beta$$\uparrow$ & $SR$$\uparrow$ & $\beta$$\uparrow$ & $SR$$\uparrow$ & $\beta$$\uparrow$ & $SR$$\uparrow$ & $\beta$$\uparrow$ \\
				\midrule
				
				\multirow{6}{*}{\rotatebox[origin=c]{90}{\shortstack{LLaMA-3.1 \\ Instruct 8B}}}
				& SPS      & 0.49 & 0.24 & 0.54 & 0.27 & 0.43 & 0.21 & 0.49 & 0.24 & 0.86 & 0.17 & 0.92 & 0.18 & 0.75 & 0.17 & 0.84 & 0.17 \\
				& EAGLE     & 0.70 & 0.35 & 0.98 & 0.37 & 0.78 & 0.35 & 0.82 & 0.36 & 1.15 & 0.20 & 1.58 & 0.30 & 1.48 & 0.26 & 1.41 & 0.25 \\
				& EAGLE-2   & 1.01 & 0.47 & 1.34 & 0.54 & 1.08 & 0.48 & 1.14 & 0.50 & 1.43 & 0.27 & 2.12 & 0.42 & 1.82 & 0.33 & 1.79 & 0.34 \\
				& GRIFFIN   & 1.17 & 0.55 & 1.50 & 0.68 & 1.20 & 0.59 & 1.29 & 0.60 & 1.58 & 0.33 & 2.61 & 0.52 & 2.01 & 0.41 & 2.07 & 0.42 \\
				\cmidrule(lr){2-18}
				& EAGLE-3        & 1.35 & 0.67 & 1.44 & 0.74 & 1.28 & 0.68 & 1.36 & 0.70 & 1.65 & 0.43 & 2.45 & 0.51 & 2.22 & 0.48 & 2.11 & 0.47 \\
				& \textbf{EAGLE-3+SMART} & \textbf{1.56} & \textbf{0.67} & \textbf{1.71} & \textbf{0.74} & \textbf{1.51} & \textbf{0.68} & $\mathbf{1.59_{+16.9\%}}$ & $\mathbf{0.80_{+14.2\%}}$ & \textbf{1.84} & \textbf{0.47} & \textbf{2.73} & \textbf{0.57} & \textbf{2.56} & \textbf{0.52} & $\mathbf{2.38_{+12.8\%}}$ & $\mathbf{0.52_{+10.6\%}}$ \\
				
				\midrule
				\multirow{5}{*}{\rotatebox[origin=c]{90}{\shortstack{Vicuna-1.3 \\ 13B}}}
				& SPS     & 0.48 & 0.25 & 0.53 & 0.28 & 0.42 & 0.21 & 0.48 & 0.25 & 0.40 & 0.16 & 0.43 & 0.17 & 0.35 & 0.15 & 0.39 & 0.16 \\
				& EAGLE   & 0.71 & 0.42 & 0.82 & 0.46 & 0.68 & 0.40 & 0.74 & 0.43 & 0.55 & 0.27 & 0.63 & 0.30 & 0.56 & 0.30 & 0.58 & 0.29 \\
				& EAGLE-2 & 0.95 & 0.54 & 1.19 & 0.60 & 0.95 & 0.52 & 1.03 & 0.55 & 0.88 & 0.38 & 0.97 & 0.42 & 0.74 & 0.38 & 0.81 & 0.40 \\
				\cmidrule(lr){2-18}
				& EAGLE-3        & 1.20 & 0.74 & 1.39 & 0.85 & 1.24 & 0.71 & 1.28 & 0.76 & 0.95 & 0.49 & 1.20 & 0.56 & 1.03 & 0.50 & 1.06 & 0.52 \\
				& \textbf{EAGLE-3+SMART} & \textbf{1.43} & \textbf{0.79} & \textbf{1.72} & \textbf{0.88} & \textbf{1.52} & \textbf{0.76} & $\mathbf{1.56_{+21.9\%}}$ & $\mathbf{0.81_{+6.6\%}}$ & \textbf{1.14} & \textbf{0.52} & \textbf{1.43} & \textbf{0.60} & \textbf{1.24} & \textbf{0.54} & $\mathbf{1.27_{+19.8\%}}$ & $\mathbf{0.55_{+5.8\%}}$ \\
				
				\midrule
				\multirow{5}{*}{\rotatebox[origin=c]{90}{\shortstack{DeepSeek\\ R1 8B}}}
				& SPS      & 0.52 & 0.26 & 0.58 & 0.29 & 0.46 & 0.22 & 0.52 & 0.26 & 0.46 & 0.18 & 0.49 & 0.19 & 0.40 & 0.17 & 0.45 & 0.18 \\
				& GRIFFIN   & 1.06 & 0.50 & 1.32 & 0.63 & 1.42 & 0.68 & 1.27 & 0.61 & 0.92 & 0.39 & 1.14 & 0.46 & 1.36 & 0.55 & 1.14 & 0.46 \\
				\cmidrule(lr){2-18}
				& EAGLE-3        & 1.24 & 0.61 & 1.49 & 0.71 & 1.61 & 0.76 & 1.46 & 0.70 & 1.06 & 0.48 & 1.22 & 0.51 & 1.56 & 0.58 & 1.28 & 0.52 \\
				& \textbf{EAGLE-3+SMART} & \textbf{1.45} & \textbf{0.69} & \textbf{1.65} & \textbf{0.77} & \textbf{1.87} & \textbf{0.82} & $\mathbf{1.68_{+15.1\%}}$ & $\mathbf{0.76_{+8.6\%}}$ & \textbf{1.21} & \textbf{0.51} & \textbf{1.34} & \textbf{0.55} & \textbf{1.62} & \textbf{0.59} & $\mathbf{1.39_{+8.6\%}}$ & $\mathbf{0.55_{+5.8\%}}$ \\
				
				\midrule
				\multirow{4}{*}{\rotatebox[origin=c]{90}{\shortstack{LLaMA\\ 3.3 70B}}}
				& SPS      & 0.98 & 0.25 & 1.09 & 0.28 & 0.86 & 0.21 & 0.98 & 0.25 & 0.81 & 0.17 & 0.86 & 0.18 & 0.70 & 0.16 & 0.79 & 0.17 \\
				\cmidrule(lr){2-18}
				& EAGLE-3        & 2.46 & 0.63 & 2.92 & 0.72 & 2.67 & 0.66 & 2.69 & 0.67 & 2.07 & 0.43 & 2.80 & 0.65 & 2.55 & 0.55 & 2.48 & 0.54 \\
				& \textbf{EAGLE-3+SMART} & \textbf{2.97} & \textbf{0.65} & \textbf{3.72} & \textbf{0.77} & \textbf{3.32} & \textbf{0.70} & $\mathbf{3.35_{+24.5\%}}$ & $\mathbf{0.70_{+4.5\%}}$ & \textbf{2.56} & \textbf{0.51} & \textbf{3.36} & \textbf{0.68} & \textbf{3.02} & \textbf{0.60} & $\mathbf{2.99_{+20.2\%}}$ & $\mathbf{0.60_{+11.1\%}}$ \\
				
				\bottomrule
			\end{tabular}%
		}
	\end{table*}

	\subsection{Main results}
	
	
	\noindent\textbf{Results on MLLMs.} Table~\ref{tab:mllm_speedup} shows that on multimodal benchmarks, SMART can substantially improve both the speedup ratio and the acceptance rate. For example, MSD+SMART consistently outperforms MSD, with average $SR$ gains of +20.2\% at $T{=}0$ and +19.9\% at $T{=}1$ across three MLLMs. The largest improvements appear on \textsc{LLaVA-1.5}: at $T{=}0$, the average $SR$ increases from $1.18\times$ to $1.53\times$ on \textsc{LLaVA-1.5-7B} and from $1.26\times$ to $1.53\times$ on \textsc{LLaVA-1.5-13B}. These gains are accompanied by higher acceptance rates: SMART preferentially expands nodes with higher expected acceptance, prunes low-benefit branches, and terminates early when no promising tokens remain. In contrast, MSD always selects a fixed number of draft tokens, even when additional tokens are unlikely to be accepted. SMART allocates the computation budget to more promising tokens and thereby improves verification efficiency.
	
	\noindent\textbf{Results on LLMs.} Table~\ref{tab:speedup} demonstrates that across four LLMs, SMART consistently improves both the speedup ratio and the acceptance rate. For instance, EAGLE-3+SMART achieves substantial speedups over EAGLE-3: +19.6\% at $T{=}0$ and +15.4\% at $T{=}1$ on average across LLMs. Improvements are consistent across chat (MT-Bench), coding (HumanEval), and math reasoning (GSM8K).  On \textsc{LLaMA-3.1-Instruct-8B}, EAGLE-3+SMART increases the average $SR$ from $1.36\times$ to $1.59\times$ at $T{=}0$ and from $2.11\times$ to $2.38\times$ at $T{=}1$. On the larger \textsc{LLaMA-3.3-70B}, SMART further raises the average $SR$ from $2.69\times$ to $3.35\times$ at $T{=}0$ and from $2.48\times$ to $2.99\times$ at $T{=}1$. Overall, SMART delivers robust gains across diverse model scales and task families. We also observe acceptance-rate improvements across all four LLMs, suggesting SMART reduces wasted draft expansions.

	\begin{table}[t]
		\centering
		\caption{Speedup comparison between MSD and SMART across different GPUs and batch sizes. SMART maintains consistent speedup while MSD degrades at large batches.}
		\label{tab:gpu-speedup}
		\resizebox{\columnwidth}{!}{%
			\begin{tabular}{c|c||ccc|c||ccc|c}
				\toprule
				\multirow{2}{*}{\textbf{GPU}} & \multirow{2}{*}{\textbf{Batch Size}} & \multicolumn{4}{c||}{\textbf{MSD}} & \multicolumn{4}{c}{\textbf{MSD + SMART (Ours)}} \\
				\cmidrule(lr){3-6} \cmidrule(lr){7-10}
				& & ChartQA & TextVQA & Hallusion & \textbf{Avg} & ChartQA & TextVQA & Hallusion & \textbf{Avg} \\
				\midrule
				\multirow{5}{*}{\rotatebox[origin=c]{90}{\shortstack{RTX Pro \\ 6000}}}
				& 1  & 2.27$\times$ & 2.09$\times$ & 2.23$\times$ & 2.20$\times$ & 2.18$\times$ & 2.10$\times$ & 2.22$\times$ & 2.17$\times$ \\
				& 8  & 1.88$\times$ & 1.80$\times$ & 1.83$\times$ & 1.84$\times$ & 1.98$\times$ & 1.85$\times$ & 1.96$\times$ & 1.93$\times$ \\
				& 16 & 1.14$\times$ & 1.26$\times$ & 1.26$\times$ & 1.22$\times$ & 1.59$\times$ & 1.55$\times$ & 1.61$\times$ & 1.58$\times$ \\
				& 24 & 0.98$\times$ & 0.95$\times$ & 0.96$\times$ & 0.96$\times$ & 1.51$\times$ & 1.41$\times$ & 1.44$\times$ & 1.45$\times$ \\
				& 32 & 0.86$\times$ & 0.79$\times$ & 0.81$\times$ & 0.82$\times$ & 1.40$\times$ & 1.38$\times$ & 1.39$\times$ & 1.39$\times$ \\
				\midrule
				\multirow{4}{*}{\rotatebox[origin=c]{90}{L40S}}
				& 1  & 1.85$\times$ & 1.79$\times$ & 1.82$\times$ & 1.82$\times$ & 1.78$\times$ & 1.76$\times$ & 1.77$\times$ & 1.77$\times$ \\
				& 4  & 1.65$\times$ & 1.58$\times$ & 1.67$\times$ & 1.63$\times$ & 1.67$\times$ & 1.59$\times$ & 1.68$\times$ & 1.65$\times$ \\
				& 8  & 1.25$\times$ & 1.20$\times$ & 1.21$\times$ & 1.22$\times$ & 1.52$\times$ & 1.44$\times$ & 1.53$\times$ & 1.50$\times$ \\
				& 12 & 0.91$\times$ & 0.89$\times$ & 0.90$\times$ & 0.90$\times$ & 1.40$\times$ & 1.37$\times$ & 1.42$\times$ & 1.40$\times$ \\
				\bottomrule
			\end{tabular}%
		}
	\end{table}

	\begin{table}[t]
		\centering
		\caption{Speedup across different token budgets on RTX Pro 6000 with batch size of 16. Budget of 200 achieves optimal performance, balancing tree size and verification cost. Lower budgets (100) under-utilize parallelism while higher budgets (300-400) incur excessive verification overhead.}
		\label{tab:budget-ablation}
		\resizebox{\columnwidth}{!}{%
			\begin{tabular}{c|ccc|c||ccc|c}
				\toprule
				\multirow{2}{*}{\textbf{Token Budget}} 
				& \multicolumn{4}{c||}{$\mathbf{T=0}$} 
				& \multicolumn{4}{c}{$\mathbf{T=1}$} \\
				\cmidrule(lr){2-5}\cmidrule(lr){6-9}
				& \textbf{ChartQA} & \textbf{TextVQA} & \textbf{Hallusion} & \textbf{Avg}
				& \textbf{ChartQA} & \textbf{TextVQA} & \textbf{Hallusion} & \textbf{Avg} \\
				\midrule
				100 & 1.39$\times$ & 1.42$\times$ & 1.47$\times$ & 1.43$\times$
				& 1.95$\times$ & 1.33$\times$ & 2.05$\times$ & 2.06$\times$ \\
				\rowcolor{gray!20}
				200 & 1.60$\times$ & 1.56$\times$ & 1.57$\times$ & \textbf{1.58$\times$}
				& 2.24$\times$ & 1.46$\times$ & 2.19$\times$ & \textbf{2.28$\times$} \\
				300 & 1.33$\times$ & 1.24$\times$ & 1.27$\times$ & 1.28$\times$
				& 1.86$\times$ & 1.16$\times$ & 1.77$\times$ & 1.85$\times$ \\
				400 & 1.32$\times$ & 1.24$\times$ & 1.26$\times$ & 1.27$\times$
				& 1.85$\times$ & 1.16$\times$ & 1.76$\times$ & 1.83$\times$ \\
				\bottomrule
			\end{tabular}%
		}
	\end{table}
	
	\begin{table}[t]
		\centering
		\begin{minipage}{0.95\columnwidth}
			\raggedright
			\caption{Speedup ablation over discount factor $\alpha$ on RTX Pro 6000 at batch size 16. $\alpha \in [0.7, 0.9]$ achieves optimal performance, balancing conservative pruning with sufficient tree expansion.}
			\label{tab:alpha-ablation}
		\end{minipage}
		
		\resizebox{0.95\columnwidth}{!}{%
			\begin{tabular}{c||ccc|c||ccc|c}
				\toprule
				\multirow{2}{*}{\textbf{$\alpha$}} & \multicolumn{4}{c||}{\textbf{Temperature = 0}} & \multicolumn{4}{c}{\textbf{Temperature = 1}} \\
				\cmidrule(lr){2-5} \cmidrule(lr){6-9}
				& ChartQA & TextVQA & Hallusion & Avg & ChartQA & TextVQA & Hallusion & Avg \\
				\midrule
				1.0 & 1.57$\times$ & 1.46$\times$ & 1.51$\times$ & 1.51$\times$ & 1.45$\times$ & 1.38$\times$ & 1.42$\times$ & 1.42$\times$ \\
				0.9 & 1.58$\times$ & 1.51$\times$ & 1.57$\times$ & 1.55$\times$ & 1.52$\times$ & 1.44$\times$ & 1.48$\times$ & 1.48$\times$ \\
				\rowcolor{gray!20}
				0.8 & 1.58$\times$ & 1.52$\times$ & 1.57$\times$ & \textbf{1.56$\times$} & 1.53$\times$ & 1.45$\times$ & 1.49$\times$ & \textbf{1.49$\times$} \\
				0.7 & 1.58$\times$ & 1.50$\times$ & 1.57$\times$ & 1.55$\times$ & 1.51$\times$ & 1.43$\times$ & 1.47$\times$ & 1.47$\times$ \\
				0.6 & 1.57$\times$ & 1.50$\times$ & 1.56$\times$ & 1.54$\times$ & 1.49$\times$ & 1.41$\times$ & 1.45$\times$ & 1.45$\times$ \\
				0.5 & 1.57$\times$ & 1.50$\times$ & 1.55$\times$ & 1.54$\times$ & 1.46$\times$ & 1.38$\times$ & 1.43$\times$ & 1.42$\times$ \\
				\bottomrule
			\end{tabular}%
		}
	\end{table}
	
\subsection{Ablation Study}
	\label{sec:ablations}
	
	\textbf{Scaling with batch size on different hardwares.}
	Table~\ref{tab:gpu-speedup} shows that SMART provides limited gains at small batch sizes, where decoding is largely memory-bound and likelihood-maximizing trees already contain most of the ``useful'' draft tokens. In this regime, SMART mainly prunes low-utility tokens but does not create additional high-utility candidates beyond those already present, leading to near ties with MSD (e.g., RTX Pro 6000 at batch $1$: $2.17\times$ vs.\ $2.20\times$). As batch size increases, execution becomes increasingly compute-bound and verification overhead grows super-linearly, making every selected draft token expensive. In this setting, MSD degrades sharply (RTX Pro 6000: $2.20\times \rightarrow 0.82\times$ from batch $1$ to $32$; L40S: $1.82\times \rightarrow 0.90\times$ from batch $1$ to $12$), whereas SMART remains robust by allocating the budget to truly beneficial tokens and avoiding wasteful verification. Consequently, SMART maintains $>1\times$ speedup throughout, retaining $1.58\times$ at batch $16$ and $1.39\times$ at batch $32$ on RTX Pro 6000, and $1.50\times$ at batch $8$ and $1.40\times$ at batch $12$ on L40S.
	
	\noindent\textbf{Verification token budget.}
	Table~\ref{tab:budget-ablation} ablates the verification token budget on RTX Pro 6000 with batch size of 16. Performance exhibits a clear optimum at budget $200$, which yields the best average speedup ($1.58\times$ at $T{=}0$ and $2.28\times$ at $T{=}1$). Lower budget ($100$) reduces speedup ($1.43\times$ at $T{=}0$), indicating under-utilization of available parallelism due to overly aggressive pruning. Larger budgets ($300$--$400$) significantly hurt speedup ($\approx 1.27$--$1.28\times$ at $T{=}0$), as additional drafted tokens increase verification work and dominate end-to-end latency. Overall, these results support allocating a moderate per-sequence verification budget that balances tree expansion against verification overhead.
	
	\noindent\textbf{Discount factor $\alpha$.}
	Table~\ref{tab:alpha-ablation} ablates the discount factor $\alpha$ used to conservatively down-weight predicted marginal benefit under draft--target mismatch. Across both temperatures, performance remains stable over a wide range of moderate discounts: $\alpha\in[0.7,0.9]$ yields the best or near-best average speedup, peaking at $\alpha=0.8$ ($1.56\times$ at $T{=}0$ and $1.49\times$ at $T{=}1$). Setting $\alpha$ too large (e.g., $\alpha=1.0$) is overly permissive, leading to aggressive expansion and higher verification overhead, which reduces speedup ($1.51\times$ at $T{=}0$ and $1.42\times$ at $T{=}1$). Conversely, smaller $\alpha$ values prune more aggressively and may discard useful draft tokens. Overall, SMART is insensitive within a wide operating range, and we use $\alpha=0.8$ by default.

\section{Conclusion}
We presented \textbf{SMART}, a training-free, system-aware framework for speedup-maximizing draft tree construction in speculative decoding. Motivated by the efficiency paradox that large draft trees can incur super-linear drafting and verification overhead, SMART casts tree growth as a sequential decision problem. By expanding nodes only when their marginal benefit--cost ratio exceeds the tree-level speedup under a per-sequence budget, SMART reduces wasteful drafting and verification and maintains robust wall-clock gains across compute-bound batching regimes and diverse GPU architectures. Across LLMs and MLLMs, SMART consistently improves strong tree-based backbones while preserving the lossless guarantee, delivering average additional speedups of \textbf{20.0\%} (MLLMs) and \textbf{15.4\%} (LLMs) without performance degradation.

\noindent{\textbf{Limitation Discussion.}}	
Due to limited compute resources, we only evaluate SMART on RTX Pro 6000 and L40S GPUs, and do not include other data-center accelerators such as A100, H100, or H200. Nonetheless, SMART is system-aware and hardware-agnostic by design, and we expect similar speedup trends on these architectures.

\section{Acknowledgement}
This work was supported by the Singapore Ministry of Education (MOE) Academic Research Fund (AcRF) Tier 1 grant (Proposal ID: 25-SIS-SMU-003).

%
%

\bibliographystyle{splncs04}
\bibliography{main}

\clearpage
\appendix 
\setcounter{table}{5}
\setcounter{page}{1}

\noindent In this supplementary material, we present more details, experiments and discussions that are not covered in the main text. This supplementary material is organized as follows:

\begin{itemize}
    \item We summarize the full SMART algorithm in Algorithm~\ref{alg:smart}.
    \item We provide additional experimental results on both standard MLLM benchmarks and LLM benchmarks under different temperature settings (Sec.~\ref{sec:supp-experiments}). 
    \item We evaluate hardware generalizability by benchmarking SMART on A100 40\,GB and H200 GPUs (Sec.~\ref{sec:supp-hardware}).
    \item We ablate the effect of tree size by sweeping fixed node-expansion width $k$ and draft depth $d$ (Sec.~\ref{sec:supp-static-trees}).
\end{itemize}

\begin{table*}[t]
    \centering
    \caption{Comparison of speedup ratio $SR$ and acceptance rate $\beta$ on standard MLLM benchmarks. The subscripts denote the relative improvement compared to the corresponding baseline. Overall, SMART yields an average $SR$ gain of \textbf{+20.0\%} across all method--model--benchmark combinations at both temperatures.}
    \label{tab:mllm_speedup_appendix}
    \resizebox{\textwidth}{!}{%
        \begin{tabular}{c c c || cc cc cc cc cc cc | ll}
            \toprule
            & & & \multicolumn{2}{c}{VQAv2} & \multicolumn{2}{c}{AI2D} & \multicolumn{2}{c}{SQA Image} & \multicolumn{2}{c}{ChartQA} & \multicolumn{2}{c}{TextVQA} & \multicolumn{2}{c}{Hallusion} & \multicolumn{2}{c}{Average} \\
            \cmidrule(lr){4-5}\cmidrule(lr){6-7}\cmidrule(lr){8-9}\cmidrule(lr){10-11}\cmidrule(lr){12-13}\cmidrule(lr){14-15}\cmidrule(lr){16-17}
            & \textbf{Model} & \textbf{Method} & $SR$$\uparrow$ & $\beta$$\uparrow$ & $SR$$\uparrow$ & $\beta$$\uparrow$ & $SR$$\uparrow$ & $\beta$$\uparrow$ & $SR$$\uparrow$ & $\beta$$\uparrow$ & $SR$$\uparrow$ & $\beta$$\uparrow$ & $SR$$\uparrow$ & $\beta$$\uparrow$ & $SR$$\uparrow$ & $\beta$$\uparrow$ \\
            \midrule

            \multirow{18}{*}{\rotatebox[origin=c]{90}{Temperature = 0}}
            & \multirow{6}{*}{\shortstack{LLaVA-1.5\\7B}}
            & Medusa  & 0.88 & 0.48 & 0.82 & 0.44 & 0.86 & 0.46 & 0.91 & 0.51 & 0.95 & 0.54 & 1.01 & 0.58 & 0.91 & 0.50 \\
            & & EAGLE-1 & 0.95 & 0.52 & 0.88 & 0.48 & 0.92 & 0.50 & 0.98 & 0.55 & 1.02 & 0.58 & 1.08 & 0.62 & 0.97 & 0.54 \\
            \cmidrule(lr){3-17}
            & & EAGLE-2 & 1.08 & 0.59 & 0.98 & 0.53 & 1.02 & 0.54 & 1.06 & 0.62 & 1.15 & 0.64 & 1.18 & 0.68 & 1.08 & 0.60 \\
            & & \textbf{EAGLE-2+SMART} & \textbf{1.37} & \textbf{0.68} & \textbf{1.24} & \textbf{0.66} & \textbf{1.31} & \textbf{0.69} & \textbf{1.36} & \textbf{0.71} & \textbf{1.47} & \textbf{0.74} & \textbf{1.51} & \textbf{0.78} & $\mathbf{1.38_{+27.8\%}}$ & $\mathbf{0.71_{+18.3\%}}$ \\
            \cmidrule(lr){3-17}
            & & MSD     & 1.23 & 0.67 & 1.09 & 0.58 & 1.09 & 0.56 & 1.14 & 0.68 & 1.26 & 0.68 & 1.26 & 0.71 & 1.18 & 0.65 \\
            & & \textbf{MSD+SMART} & \textbf{1.55} & \textbf{0.81} & \textbf{1.45} & \textbf{0.74} & \textbf{1.44} & \textbf{0.72} & \textbf{1.59} & \textbf{0.82} & \textbf{1.55} & \textbf{0.79} & \textbf{1.62} & \textbf{0.83} & $\mathbf{1.53_{+29.7\%}}$ & $\mathbf{0.79_{+21.5\%}}$ \\
            \cmidrule(lr){2-17}
            & \multirow{6}{*}{\shortstack{LLaVA-1.5\\13B}}
            & Medusa  & 0.95 & 0.44 & 0.88 & 0.41 & 1.01 & 0.46 & 1.10 & 0.54 & 0.91 & 0.42 & 0.88 & 0.44 & 0.96 & 0.45 \\
            & & EAGLE-1 & 1.02 & 0.48 & 0.95 & 0.45 & 1.08 & 0.50 & 1.18 & 0.58 & 0.98 & 0.46 & 0.95 & 0.48 & 1.03 & 0.49 \\
            \cmidrule(lr){3-17}
            & & EAGLE-2 & 1.15 & 0.54 & 1.08 & 0.50 & 1.18 & 0.54 & 1.32 & 0.62 & 1.12 & 0.51 & 1.08 & 0.51 & 1.16 & 0.54 \\
            & & \textbf{EAGLE-2+SMART} & \textbf{1.39} & \textbf{0.68} & \textbf{1.30} & \textbf{0.64} & \textbf{1.42} & \textbf{0.69} & \textbf{1.59} & \textbf{0.76} & \textbf{1.35} & \textbf{0.66} & \textbf{1.31} & \textbf{0.65} & $\mathbf{1.40_{+20.7\%}}$ & $\mathbf{0.68_{+25.9\%}}$ \\
            \cmidrule(lr){3-17}
            & & MSD     & 1.30 & 0.59 & 1.17 & 0.53 & 1.28 & 0.57 & 1.45 & 0.66 & 1.22 & 0.54 & 1.17 & 0.54 & 1.26 & 0.57 \\
            & & \textbf{MSD+SMART} & \textbf{1.56} & \textbf{0.76} & \textbf{1.42} & \textbf{0.71} & \textbf{1.53} & \textbf{0.76} & \textbf{1.72} & \textbf{0.84} & \textbf{1.51} & \textbf{0.73} & \textbf{1.42} & \textbf{0.72} & $\mathbf{1.53_{+21.4\%}}$ & $\mathbf{0.75_{+31.6\%}}$ \\
            \cmidrule(lr){2-17}
            & \multirow{6}{*}{\shortstack{Qwen2VL\\7B Instruct}}
            & Medusa  & 0.85 & 0.54 & 0.79 & 0.48 & 0.82 & 0.47 & 0.98 & 0.54 & 0.88 & 0.50 & 0.91 & 0.54 & 0.87 & 0.51 \\
            & & EAGLE-1 & 0.92 & 0.58 & 0.85 & 0.52 & 0.88 & 0.51 & 1.05 & 0.58 & 0.95 & 0.54 & 0.98 & 0.58 & 0.94 & 0.55 \\
            \cmidrule(lr){3-17}
            & & EAGLE-2 & 1.05 & 0.65 & 0.96 & 0.56 & 0.98 & 0.55 & 1.15 & 0.62 & 1.08 & 0.58 & 1.08 & 0.61 & 1.05 & 0.60 \\
            & & \textbf{EAGLE-2+SMART} & \textbf{1.16} & \textbf{0.77} & \textbf{1.06} & \textbf{0.71} & \textbf{1.09} & \textbf{0.69} & \textbf{1.27} & \textbf{0.77} & \textbf{1.19} & \textbf{0.56} & \textbf{1.19} & \textbf{0.76} & $\mathbf{1.16_{+10.5\%}}$ & $\mathbf{0.71_{+18.3\%}}$ \\
            \cmidrule(lr){3-17}
            & & MSD     & 1.18 & 0.71 & 1.05 & 0.58 & 1.06 & 0.57 & 1.24 & 0.65 & 1.16 & 0.60 & 1.15 & 0.63 & 1.14 & 0.62 \\
            & & \textbf{MSD+SMART} & \textbf{1.22} & \textbf{0.86} & \textbf{1.24} & \textbf{0.77} & \textbf{1.26} & \textbf{0.74} & \textbf{1.34} & \textbf{0.84} & \textbf{1.24} & \textbf{0.57} & \textbf{1.22} & \textbf{0.82} & $\mathbf{1.25_{+9.6\%}}$ & $\mathbf{0.77_{+24.2\%}}$ \\

            \midrule

            \multirow{18}{*}{\rotatebox[origin=c]{90}{Temperature = 1}}
            & \multirow{6}{*}{\shortstack{LLaVA-1.5\\7B}}
            & Medusa  & 1.74 & 0.24 & 1.43 & 0.28 & 1.58 & 0.29 & 1.33 & 0.26 & 0.97 & 0.19 & 1.46 & 0.26 & 1.42 & 0.25 \\
            & & EAGLE-1 & 1.85 & 0.26 & 1.52 & 0.30 & 1.68 & 0.31 & 1.42 & 0.28 & 1.05 & 0.21 & 1.55 & 0.28 & 1.51 & 0.27 \\
            \cmidrule(lr){3-17}
            & & EAGLE-2 & 2.05 & 0.28 & 1.65 & 0.32 & 1.85 & 0.33 & 1.52 & 0.29 & 1.15 & 0.23 & 1.68 & 0.30 & 1.65 & 0.29 \\
            & & \textbf{EAGLE-2+SMART} & \textbf{2.63} & \textbf{0.36} & \textbf{2.12} & \textbf{0.41} & \textbf{2.37} & \textbf{0.43} & \textbf{1.95} & \textbf{0.37} & \textbf{1.47} & \textbf{0.30} & \textbf{2.14} & \textbf{0.39} & $\mathbf{2.11_{+28.5\%}}$ & $\mathbf{0.38_{+31.0\%}}$ \\
            \cmidrule(lr){3-17}
            & & MSD     & 2.21 & 0.30 & 1.77 & 0.34 & 2.00 & 0.35 & 1.63 & 0.31 & 1.22 & 0.24 & 1.77 & 0.31 & 1.77 & 0.31 \\
            & & \textbf{MSD+SMART} & \textbf{2.84} & \textbf{0.40} & \textbf{2.41} & \textbf{0.45} & \textbf{2.56} & \textbf{0.46} & \textbf{2.24} & \textbf{0.43} & \textbf{1.46} & \textbf{0.33} & \textbf{2.19} & \textbf{0.41} & $\mathbf{2.28_{+28.8\%}}$ & $\mathbf{0.42_{+35.5\%}}$ \\
            \cmidrule(lr){2-17}
            & \multirow{6}{*}{\shortstack{LLaVA-1.5\\13B}}
            & Medusa  & 1.67 & 0.27 & 1.39 & 0.30 & 1.91 & 0.33 & 1.36 & 0.26 & 1.07 & 0.20 & 1.23 & 0.24 & 1.44 & 0.27 \\
            & & EAGLE-1 & 1.78 & 0.29 & 1.48 & 0.32 & 2.02 & 0.35 & 1.45 & 0.28 & 1.15 & 0.22 & 1.32 & 0.26 & 1.53 & 0.29 \\
            \cmidrule(lr){3-17}
            & & EAGLE-2 & 1.95 & 0.31 & 1.58 & 0.33 & 2.25 & 0.36 & 1.58 & 0.29 & 1.25 & 0.24 & 1.42 & 0.27 & 1.67 & 0.30 \\
            & & \textbf{EAGLE-2+SMART} & \textbf{2.37} & \textbf{0.38} & \textbf{1.91} & \textbf{0.40} & \textbf{2.73} & \textbf{0.44} & \textbf{1.93} & \textbf{0.35} & \textbf{1.52} & \textbf{0.29} & \textbf{1.72} & \textbf{0.34} & $\mathbf{2.03_{+21.6\%}}$ & $\mathbf{0.37_{+23.3\%}}$ \\
            \cmidrule(lr){3-17}
            & & MSD     & 2.14 & 0.33 & 1.71 & 0.35 & 2.40 & 0.38 & 1.71 & 0.31 & 1.33 & 0.25 & 1.48 & 0.28 & 1.79 & 0.31 \\
            & & \textbf{MSD+SMART} & \textbf{2.54} & \textbf{0.42} & \textbf{2.23} & \textbf{0.47} & \textbf{2.92} & \textbf{0.46} & \textbf{1.98} & \textbf{0.40} & \textbf{1.63} & \textbf{0.34} & \textbf{1.78} & \textbf{0.40} & $\mathbf{2.18_{+21.8\%}}$ & $\mathbf{0.41_{+32.3\%}}$ \\
            \cmidrule(lr){2-17}
            & \multirow{6}{*}{\shortstack{Qwen2VL\\7B Instruct}}
            & Medusa  & 1.24 & 0.52 & 0.91 & 0.38 & 1.49 & 0.42 & 0.95 & 0.42 & 1.14 & 0.42 & 0.88 & 0.42 & 1.10 & 0.43 \\
            & & EAGLE-1 & 1.32 & 0.55 & 0.98 & 0.41 & 1.58 & 0.45 & 1.02 & 0.45 & 1.22 & 0.45 & 0.95 & 0.45 & 1.18 & 0.46 \\
            \cmidrule(lr){3-17}
            & & EAGLE-2 & 1.42 & 0.59 & 1.06 & 0.43 & 1.68 & 0.47 & 1.08 & 0.46 & 1.32 & 0.47 & 1.02 & 0.47 & 1.26 & 0.48 \\
            & & \textbf{EAGLE-2+SMART} & \textbf{1.55} & \textbf{0.64} & \textbf{1.17} & \textbf{0.47} & \textbf{1.84} & \textbf{0.52} & \textbf{1.18} & \textbf{0.51} & \textbf{1.44} & \textbf{0.52} & \textbf{1.11} & \textbf{0.52} & $\mathbf{1.38_{+9.5\%}}$ & $\mathbf{0.53_{+10.4\%}}$ \\
            \cmidrule(lr){3-17}
            & & MSD     & 1.50 & 0.62 & 1.12 & 0.44 & 1.79 & 0.48 & 1.15 & 0.48 & 1.38 & 0.48 & 1.07 & 0.48 & 1.33 & 0.50 \\
            & & \textbf{MSD+SMART} & \textbf{1.69} & \textbf{0.65} & \textbf{1.25} & \textbf{0.53} & \textbf{2.04} & \textbf{0.55} & \textbf{1.16} & \textbf{0.58} & \textbf{1.45} & \textbf{0.56} & \textbf{1.10} & \textbf{0.50} & $\mathbf{1.45_{+9.0\%}}$ & $\mathbf{0.56_{+12.0\%}}$ \\

            \bottomrule
        \end{tabular}%
    }
\end{table*}

\begin{table*}[t]
    \centering
    \caption{
        Comparison of speedup ratio ($SR$) and acceptance rate ($\beta$) on standard LLM benchmarks under temperature settings $T \in \{0,1\}$. Subscripts denote the relative improvement over the corresponding baseline. Overall, SMART consistently improves all speculative decoding baselines across all models and benchmarks, yielding an average $SR$ gain of \textbf{+15.4\%} across all method–model–benchmark combinations.
    }
    \label{tab:speedup_appendix}
    \resizebox{\textwidth}{!}{%
        \begin{tabular}{cc||cccccc|ll||cccccc|ll}
            \toprule
            & & \multicolumn{8}{c||}{\textbf{Temperature = 0}} & \multicolumn{8}{c}{\textbf{Temperature = 1}} \\
            \noalign{\smallskip}
            \cmidrule(lr){3-10} \cmidrule(lr){11-18}
            \textbf{Model} & \textbf{Method} & \multicolumn{2}{c}{MT-bench} & \multicolumn{2}{c}{HumanEval} & \multicolumn{2}{c|}{GSM8K} & \multicolumn{2}{c||}{Average} & \multicolumn{2}{c}{MT-bench} & \multicolumn{2}{c}{HumanEval} & \multicolumn{2}{c|}{GSM8K} & \multicolumn{2}{c}{Average} \\
            \cmidrule(lr){3-10} \cmidrule(lr){11-18}
            & & $SR$$\uparrow$ & $\beta$$\uparrow$ & $SR$$\uparrow$ & $\beta$$\uparrow$ & $SR$$\uparrow$ & $\beta$$\uparrow$ & $SR$$\uparrow$ & $\beta$$\uparrow$ & $SR$$\uparrow$ & $\beta$$\uparrow$ & $SR$$\uparrow$ & $\beta$$\uparrow$ & $SR$$\uparrow$ & $\beta$$\uparrow$ & $SR$$\uparrow$ & $\beta$$\uparrow$ \\
            \midrule
            
            \multirow{10}{*}{\rotatebox[origin=c]{90}{\shortstack{LLaMA-3.1 \\ Instruct 8B}}}
            & SPS      & 0.49 & 0.24 & 0.54 & 0.27 & 0.43 & 0.21 & 0.49 & 0.24 & 0.86 & 0.17 & 0.92 & 0.18 & 0.75 & 0.17 & 0.84 & 0.17 \\
            & EAGLE     & 0.70 & 0.35 & 0.98 & 0.37 & 0.78 & 0.35 & 0.82 & 0.36 & 1.15 & 0.20 & 1.58 & 0.30 & 1.48 & 0.26 & 1.41 & 0.25 \\
            \cmidrule(lr){2-18}
            & EAGLE-2   & 1.01 & 0.47 & 1.34 & 0.54 & 1.08 & 0.48 & 1.14 & 0.50 & 1.43 & 0.27 & 2.12 & 0.42 & 1.82 & 0.33 & 1.79 & 0.34 \\
            & \textbf{EAGLE-2+SMART}
              & \textbf{1.16} & \textbf{0.53} & \textbf{1.54} & \textbf{0.60} & \textbf{1.24} & \textbf{0.54}
              & $\mathbf{1.31_{+14.9\%}}$ & $\mathbf{0.56_{+12.6\%}}$
              & \textbf{1.59} & \textbf{0.29} & \textbf{2.35} & \textbf{0.46} & \textbf{2.02} & \textbf{0.36}
              & $\mathbf{1.99_{+11.2\%}}$ & $\mathbf{0.37_{+9.1\%}}$ \\
            \cmidrule(lr){2-18}
            & GRIFFIN   & 1.17 & 0.55 & 1.50 & 0.68 & 1.20 & 0.59 & 1.29 & 0.60 & 1.58 & 0.33 & 2.61 & 0.52 & 2.01 & 0.41 & 2.07 & 0.42 \\
            & \textbf{GRIFFIN+SMART}
              & \textbf{1.38} & \textbf{0.61} & \textbf{1.77} & \textbf{0.75} & \textbf{1.42} & \textbf{0.65}
              & $\mathbf{1.52_{+17.8\%}}$ & $\mathbf{0.66_{+9.5\%}}$
              & \textbf{1.79} & \textbf{0.36} & \textbf{2.96} & \textbf{0.56} & \textbf{2.28} & \textbf{0.44}
              & $\mathbf{2.33_{+12.6\%}}$ & $\mathbf{0.45_{+8.1\%}}$ \\
            \cmidrule(lr){2-18}
            & GTO            & 1.39 & 0.68 & 1.48 & 0.75 & 1.32 & 0.69 & 1.40 & 0.71 & 1.70 & 0.44 & 2.52 & 0.52 & 2.29 & 0.49 & 2.17 & 0.48 \\
            & \textbf{GTO+SMART}
              & \textbf{1.58} & \textbf{0.75} & \textbf{1.69} & \textbf{0.83} & \textbf{1.50} & \textbf{0.77}
              & $\mathbf{1.60_{+14.3\%}}$ & $\mathbf{0.79_{+11.4\%}}$
              & \textbf{1.88} & \textbf{0.47} & \textbf{2.78} & \textbf{0.56} & \textbf{2.53} & \textbf{0.53}
              & $\mathbf{2.40_{+10.6\%}}$ & $\mathbf{0.52_{+8.1\%}}$ \\
              \cmidrule(lr){2-18}
            & DFLASH         & 1.09 & 0.60 & 1.60 & 0.83 & 1.48 & 0.75 & 1.39 & 0.73 & 0.98 & 0.88 & 1.33 & 0.99 & 1.20 & 0.98 & 1.17 & 0.95 \\
            & \textbf{DFLASH+SMART}
              & \textbf{1.24} & \textbf{0.74} & \textbf{1.80} & \textbf{0.85} & \textbf{1.77} & \textbf{0.81}
              & $\mathbf{1.60_{+15.3\%}}$ & $\mathbf{0.80_{+10.1\%}}$
              & \textbf{1.08} & \textbf{0.92} & \textbf{1.56} & \textbf{1.00} & \textbf{1.44} & \textbf{0.98}
              & $\mathbf{1.36_{+16.2\%}}$ & $\mathbf{0.97_{+1.8\%}}$ \\
            \cmidrule(lr){2-18}
            & EAGLE-3        & 1.35 & 0.67 & 1.44 & 0.74 & 1.28 & 0.68 & 1.36 & 0.70 & 1.65 & 0.43 & 2.45 & 0.51 & 2.22 & 0.48 & 2.11 & 0.47 \\
            & \textbf{EAGLE-3+SMART} & \textbf{1.56} & \textbf{0.67} & \textbf{1.71} & \textbf{0.74} & \textbf{1.51} & \textbf{0.68} & $\mathbf{1.59_{+16.9\%}}$ & $\mathbf{0.80_{+14.2\%}}$ & \textbf{1.84} & \textbf{0.47} & \textbf{2.73} & \textbf{0.57} & \textbf{2.56} & \textbf{0.52} & $\mathbf{2.38_{+12.8\%}}$ & $\mathbf{0.52_{+10.6\%}}$ \\
            
            \midrule
            \multirow{8}{*}{\rotatebox[origin=c]{90}{\shortstack{Vicuna-1.3 \\ 13B}}}
            & SPS     & 0.48 & 0.25 & 0.53 & 0.28 & 0.42 & 0.21 & 0.48 & 0.25 & 0.40 & 0.16 & 0.43 & 0.17 & 0.35 & 0.15 & 0.39 & 0.16 \\
            & EAGLE   & 0.71 & 0.42 & 0.82 & 0.46 & 0.68 & 0.40 & 0.74 & 0.43 & 0.55 & 0.27 & 0.63 & 0.30 & 0.56 & 0.30 & 0.58 & 0.29 \\
            \cmidrule(lr){2-18}
            & EAGLE-2 & 0.95 & 0.54 & 1.19 & 0.60 & 0.95 & 0.52 & 1.03 & 0.55 & 0.88 & 0.38 & 0.97 & 0.42 & 0.74 & 0.38 & 0.81 & 0.40 \\
            & \textbf{EAGLE-2+SMART}
              & \textbf{1.14} & \textbf{0.58} & \textbf{1.43} & \textbf{0.64} & \textbf{1.14} & \textbf{0.56}
              & $\mathbf{1.24_{+20.4\%}}$ & $\mathbf{0.59_{+7.5\%}}$
              & \textbf{1.04} & \textbf{0.40} & \textbf{1.14} & \textbf{0.45} & \textbf{0.87} & \textbf{0.40}
              & $\mathbf{0.95_{+17.3\%}}$ & $\mathbf{0.42_{+5.2\%}}$ \\
            \cmidrule(lr){2-18}
            & GTO            & 1.24 & 0.75 & 1.43 & 0.87 & 1.28 & 0.72 & 1.32 & 0.78 & 0.98 & 0.50 & 1.24 & 0.57 & 1.06 & 0.51 & 1.09 & 0.53 \\
            & \textbf{GTO+SMART}
              & \textbf{1.45} & \textbf{0.79} & \textbf{1.67} & \textbf{0.91} & \textbf{1.50} & \textbf{0.76}
              & $\mathbf{1.54_{+16.7\%}}$ & $\mathbf{0.82_{+5.3\%}}$
              & \textbf{1.13} & \textbf{0.52} & \textbf{1.43} & \textbf{0.59} & \textbf{1.22} & \textbf{0.53}
              & $\mathbf{1.25_{+14.7\%}}$ & $\mathbf{0.55_{+4.2\%}}$ \\
            \cmidrule(lr){2-18}
            & EAGLE-3        & 1.20 & 0.74 & 1.39 & 0.85 & 1.24 & 0.71 & 1.28 & 0.76 & 0.95 & 0.49 & 1.20 & 0.56 & 1.03 & 0.50 & 1.06 & 0.52 \\
            & \textbf{EAGLE-3+SMART} & \textbf{1.43} & \textbf{0.79} & \textbf{1.72} & \textbf{0.88} & \textbf{1.52} & \textbf{0.76} & $\mathbf{1.56_{+21.9\%}}$ & $\mathbf{0.81_{+6.6\%}}$ & \textbf{1.14} & \textbf{0.52} & \textbf{1.43} & \textbf{0.60} & \textbf{1.24} & \textbf{0.54} & $\mathbf{1.27_{+19.8\%}}$ & $\mathbf{0.55_{+5.8\%}}$ \\
            
            \midrule
            \multirow{7}{*}{\rotatebox[origin=c]{90}{\shortstack{DeepSeek\\ R1 8B}}}
            & SPS      & 0.52 & 0.26 & 0.58 & 0.29 & 0.46 & 0.22 & 0.52 & 0.26 & 0.46 & 0.18 & 0.49 & 0.19 & 0.40 & 0.17 & 0.45 & 0.18 \\
            \cmidrule(lr){2-18}
            & GRIFFIN   & 1.06 & 0.50 & 1.32 & 0.63 & 1.42 & 0.68 & 1.27 & 0.61 & 0.92 & 0.39 & 1.14 & 0.46 & 1.36 & 0.55 & 1.14 & 0.46 \\
            & \textbf{GRIFFIN+SMART}
              & \textbf{1.21} & \textbf{0.55} & \textbf{1.50} & \textbf{0.69} & \textbf{1.62} & \textbf{0.74}
              & $\mathbf{1.45_{+14.2\%}}$ & $\mathbf{0.66_{+8.5\%}}$
              & \textbf{1.00} & \textbf{0.41} & \textbf{1.24} & \textbf{0.49} & \textbf{1.48} & \textbf{0.58}
              & $\mathbf{1.24_{+8.8\%}}$ & $\mathbf{0.49_{+6.7\%}}$ \\
            \cmidrule(lr){2-18}
            & GTO            & 1.28 & 0.62 & 1.53 & 0.72 & 1.66 & 0.78 & 1.50 & 0.71 & 1.09 & 0.49 & 1.26 & 0.52 & 1.61 & 0.59 & 1.32 & 0.53 \\
            & \textbf{GTO+SMART}
              & \textbf{1.43} & \textbf{0.66} & \textbf{1.71} & \textbf{0.77} & \textbf{1.86} & \textbf{0.83}
              & $\mathbf{1.69_{+12.7\%}}$ & $\mathbf{0.76_{+7.2\%}}$
              & \textbf{1.17} & \textbf{0.51} & \textbf{1.35} & \textbf{0.55} & \textbf{1.72} & \textbf{0.62}
              & $\mathbf{1.41_{+6.8\%}}$ & $\mathbf{0.56_{+5.8\%}}$ \\
            \cmidrule(lr){2-18}
            & EAGLE-3        & 1.24 & 0.61 & 1.49 & 0.71 & 1.61 & 0.76 & 1.46 & 0.70 & 1.06 & 0.48 & 1.22 & 0.51 & 1.56 & 0.58 & 1.28 & 0.52 \\
            & \textbf{EAGLE-3+SMART} & \textbf{1.45} & \textbf{0.69} & \textbf{1.65} & \textbf{0.77} & \textbf{1.87} & \textbf{0.82} & $\mathbf{1.68_{+15.1\%}}$ & $\mathbf{0.76_{+8.6\%}}$ & \textbf{1.21} & \textbf{0.51} & \textbf{1.34} & \textbf{0.55} & \textbf{1.62} & \textbf{0.59} & $\mathbf{1.39_{+8.6\%}}$ & $\mathbf{0.55_{+5.8\%}}$ \\
            
            \midrule
            \multirow{5}{*}{\rotatebox[origin=c]{90}{\shortstack{LLaMA\\ 3.3 70B}}}
            & SPS      & 0.98 & 0.25 & 1.09 & 0.28 & 0.86 & 0.21 & 0.98 & 0.25 & 0.81 & 0.17 & 0.86 & 0.18 & 0.70 & 0.16 & 0.79 & 0.17 \\
            \cmidrule(lr){2-18}
            & GTO & 2.53 & 0.64 & 3.00 & 0.73 & 2.75 & 0.67 & 2.76 & 0.68 & 2.13 & 0.44 & 2.88 & 0.66 & 2.63 & 0.56 & 2.55 & 0.55 \\
            & \textbf{GTO+SMART}
              & \textbf{2.96} & \textbf{0.66} & \textbf{3.67} & \textbf{0.77} & \textbf{3.30} & \textbf{0.70}
              & $\mathbf{3.31_{+20.0\%}}$ & $\mathbf{0.71_{+4.4\%}}$
              & \textbf{2.53} & \textbf{0.50} & \textbf{3.33} & \textbf{0.68} & \textbf{3.01} & \textbf{0.60}
              & $\mathbf{2.96_{+16.1\%}}$ & $\mathbf{0.59_{+7.8\%}}$ \\
            \cmidrule(lr){2-18}
            & EAGLE-3        & 2.46 & 0.63 & 2.92 & 0.72 & 2.67 & 0.66 & 2.69 & 0.67 & 2.07 & 0.43 & 2.80 & 0.65 & 2.55 & 0.55 & 2.48 & 0.54 \\
            & \textbf{EAGLE-3+SMART} & \textbf{2.97} & \textbf{0.65} & \textbf{3.72} & \textbf{0.77} & \textbf{3.32} & \textbf{0.70} & $\mathbf{3.35_{+24.5\%}}$ & $\mathbf{0.70_{+4.5\%}}$ & \textbf{2.56} & \textbf{0.51} & \textbf{3.36} & \textbf{0.68} & \textbf{3.02} & \textbf{0.60} & $\mathbf{2.99_{+20.2\%}}$ & $\mathbf{0.60_{+11.1\%}}$ \\
            
            \bottomrule
        \end{tabular}%
    }
\end{table*}

\section{Additional Experiment Results}
\label{sec:supp-experiments}
\noindent\textbf{Additional Results on MLLMs.} 
Table~\ref{tab:mllm_speedup_appendix} provides a more comprehensive comparison between baseline speculative decoding methods and their SMART-enhanced variants on multimodal benchmarks. Similar to the analysis in the main text, SMART consistently improves both the speedup ratio ($SR$) and the acceptance rate ($\beta$) when integrated with existing frameworks. For example, EAGLE-2~\cite{li2024eagle2}+SMART consistently outperforms EAGLE-2 across all three MLLMs and all six benchmarks. On LLaVA-1.5-7B~\cite{liu2023visual} the average $SR$ increases from $1.08\times$ to $1.38\times$ at $T{=}0$ and from $1.65\times$ to $2.11\times$ at $T{=}1$, corresponding to relative improvements of $+27.8\%$ and $+28.5\%$, respectively. Similar trends are observed on LLaVA-1.5-13B, where SMART raises the average $SR$ from $1.16\times$ to $1.40\times$ at $T{=}0$ and from $1.67\times$ to $2.03\times$ at $T{=}1$. These improvements are consistently accompanied by higher acceptance rates. As discussed in the main text, SMART prioritizes tokens with higher expected acceptance probability, prunes branches with low expected benefit, and stops exploration early when no promising candidates remain. In contrast, baseline methods such as EAGLE-2 or MSD~\cite{lin2025speculative} typically expand a fixed number of draft tokens, which may lead to wasted verification when many drafted tokens are unlikely to be accepted.

\begin{algorithm}[t]
\caption{SMART: Greedy Draft-Tree Construction}
\label{alg:smart}
\begin{algorithmic}[1]
\REQUIRE Verification budget $B_{\mathrm{verify}}$, batch size $b$, top-$k$, max depth $d$, discount $\alpha$
\ENSURE Final selected node set $S_L$ (draft tree)
\STATE $B \gets B_{\mathrm{verify}}/b$ \COMMENT{per-sequence budget}
\STATE Initialize $S_0 \gets \{\mathrm{root}\}$, $A_0 \gets \{\mathrm{root}\}$
\FOR{layer $\ell = 1$ to $d$}
    \STATE $\mathcal{U}_\ell \gets \mathcal{U}_\ell(A_{\ell-1})$ \COMMENT{expand each node in $A_{\ell-1}$ with top-$k$ candidates}
    \STATE Compute global terms on current tree:
    \STATE \hspace{1em}$C_{\mathrm{target}} \gets c_T\,L^{\mathrm{tree}}(S_{\ell-1})$, \quad $C_{\mathrm{spec}} \gets C_{\mathrm{draft}}(S_{\ell-1})+C_{\mathrm{verify}}(S_{\ell-1})$
    \STATE Initialize $A_\ell \gets \emptyset$
    \FOR{each candidate $u \in \mathcal{U}_\ell$}
        \STATE Compute $\Delta C_{\mathrm{target}}(u)$ via Eqn.~\eqref{eq:marginal_benefit} and $\Delta C_{\mathrm{spec}}(u)$ via Eqn.~\eqref{eq:marginal_cost}
        \STATE $e_\ell(u) \gets \mathbf{1}\!\left[\alpha \cdot \frac{\Delta C_{\mathrm{target}}(u)}{\Delta C_{\mathrm{spec}}(u)} - \frac{C_{\mathrm{target}}}{C_{\mathrm{spec}}}> 0\right]$
        \IF{$e_\ell(u) = 1$}
            \STATE $A_\ell \gets A_\ell \cup \{u\}$
        \ENDIF
    \ENDFOR
    \STATE $S_\ell \gets S_{\ell-1} \cup A_\ell$
    \IF{$A_\ell = \emptyset$ \OR $|S_\ell| \ge B$}
        \STATE \textbf{break}
    \ENDIF
\ENDFOR
\STATE Let $L$ be the last executed layer; \textbf{return} $S_L$
\end{algorithmic}
\end{algorithm}

\noindent\textbf{Additional Results on LLMs.} 
Table~\ref{tab:speedup_appendix} reports a detailed comparison between baseline speculative decoding methods and their SMART-enhanced versions on standard LLM benchmarks. Consistent with the trends observed in the main text, SMART improves all evaluated baselines, including EAGLE-2, GRIFFIN~\cite{hu2025griffin}, GTO~\cite{hu2025bridging}, and DFLASH~\cite{chen2026dflash}. For instance, when applied to EAGLE-2 on LLaMA-3.1-Instruct-8B~\cite{grattafiori2024llama}, SMART increases the average $SR$ from $1.14\times$ to $1.31\times$ at $T{=}0$ and from $1.79\times$ to $1.99\times$ at $T{=}1$. Similarly, GRIFFIN+SMART improves the average $SR$ from $1.29\times$ to $1.52\times$ at $T{=}0$ and from $2.07\times$ to $2.33\times$ at $T{=}1$. GTO also benefits from SMART, with the average $SR$ increasing from $1.40\times$ to $1.60\times$ at $T{=}0$ and from $2.17\times$ to $2.40\times$ at $T{=}1$ on the same model. Notably, SMART also generalizes to DFLASH, a block-diffusion-based draft model that generates all draft tokens in a single non-autoregressive forward pass rather than autoregressively as in the EAGLE family. Since DFLASH produces independent top-$k$ candidates at each position, we construct an EAGLE-2 style tree by taking the Cartesian product of candidates across positions and pruning to the top-$g$ tokens by cumulative probability. Applying SMART to this tree-structured DFLASH improves the average $SR$ from $1.39\times$ to $1.60\times$ ($+15.3\%$) at $T{=}0$ and from $1.17\times$ to $1.36\times$ ($+16.2\%$) at $T{=}1$, demonstrating that SMART is effective even when the underlying draft mechanism is fundamentally different from autoregressive speculation. These results indicate that the benefits of SMART scale well with model size and remain consistent across different task domains, including dialogue generation, code generation, and mathematical reasoning, as well as across different draft model paradigms.

\noindent
Overall, these additional results further confirm the generality of SMART. Across both LLM and MLLM benchmarks, SMART consistently improves speculative decoding baselines such as EAGLE-2, GRIFFIN, GTO, DFLASH, and MSD without modifying their underlying architectures. This demonstrates that SMART can serve as a lightweight plug-in enhancement that can be readily integrated into existing speculative decoding frameworks to improve decoding efficiency.

\clearpage

\section{Additional Hardware Evaluation}
\label{sec:supp-hardware}

To evaluate hardware generalizability beyond the RTX Pro 6000 and L40S
GPUs used in the main paper, we additionally benchmark SMART on
A100 40\,GB and H200 GPUs using LLaMA-3.1-Instruct-8B with MSD as
the base speculative decoding system. Results are shown in
Table~\ref{tab:supp-hardware}.

In memory-bandwidth-bound regimes (e.g., batch size 1), the
verification overhead of a large draft tree is amortized by weight
loading, so the baseline MSD already achieves strong speedup and
SMART's pruning offers limited additional gain. However, once the
batch size exceeds a device-specific compute-saturation threshold---approximately
$b \ge 8$ on A100 and $b \ge 32$ on H200---MSD's fixed tree becomes
increasingly costly to verify, while SMART's marginal expansion rule
adapts the tree shape to the tighter hardware budget. At batch size 10
on A100, SMART improves MSD's speedup from $1.41\times$ to
$1.63\times$; at batch size 48 on H200, the gap widens from
$1.18\times$ to $1.44\times$. These results confirm that SMART's
gains are consistent across GPU architectures spanning different
memory bandwidths and compute capacities.

\begin{table}[t]
\setlength{\tabcolsep}{2.5pt}
\centering
\caption{Speedup over autoregressive decoding on A100 40\,GB and H200
  GPUs (LLaMA-3.1-Instruct-8B, MT-Bench). SMART is applied on top of
  MSD. BS\,=\,batch size.}
\label{tab:supp-hardware}
\begin{minipage}{0.49\linewidth}
\centering
\scalebox{1}{
\begin{tabular}{cc|cc}
\toprule
&   BS & MSD & +SMART \\
\midrule
\multirow{4}{*}{{\textbf{A100}}}
& 1  & 1.96$\times$ & 1.90$\times$ \\
& 4  & 1.73$\times$ & 1.75$\times$ \\
& 8  & 1.48$\times$ & 1.70$\times$ \\
& 10 & 1.41$\times$ & 1.63$\times$ \\
\bottomrule
\end{tabular}}
\end{minipage}
\hfill
\begin{minipage}{0.49\linewidth}
\centering
\scalebox{1}{
\begin{tabular}{cc|cc}
\toprule
& BS & MSD & +SMART \\
\midrule
\multirow{4}{*}{{\textbf{H200}}}
& 1  & 2.82$\times$ & 2.71$\times$ \\
& 16 & 1.95$\times$ & 2.05$\times$ \\
& 32 & 1.38$\times$ & 1.66$\times$ \\
& 48 & 1.18$\times$ & 1.44$\times$ \\
\bottomrule
\end{tabular}}
\end{minipage}
\end{table}

\begin{table*}[t]
\setlength{\tabcolsep}{12pt}
\centering
\caption{Speedup ratio of SMART over fixed-$(k,d)$ MSD
  configurations (LLaMA-3.1-Instruct-8B, MT-Bench, batch size 16, RTX
  Pro 6000). All values $>1$: SMART is faster in every case.}
\label{tab:supp-static}
\begin{tabular}{c|ccccc}
\toprule
$k \;\backslash\; d$ & 3 & 4 & 5 & 6 & 7 \\
\midrule
3  & 1.38 & 1.37 & 1.26 & 1.23 & 1.26 \\
5  & 1.32 & 1.32 & 1.24 & 1.20 & 1.23 \\
7  & 1.16 & 1.16 & 1.13 & 1.13 & 1.13 \\
10 & 1.15 & 1.16 & 1.11 & 1.09 & 1.10 \\
\bottomrule
\end{tabular}
\end{table*}

\section{Ablation on Tree Size}
\label{sec:supp-static-trees}

We ablate the effect of tree size by running MSD with fixed
node-expansion width top\nobreakdash-$k \in \{3, 5, 7, 10\}$ and
fixed draft depth $d \in \{3, 4, 5, 6, 7\}$.
Table~\ref{tab:supp-static} reports the ratio of SMART's speedup to
each fixed-$(k,d)$ configuration's speedup (values ${>}1$ indicate
SMART is faster).

SMART outperforms every fixed $(k,d)$ setting, with an average
improvement of 21\%. Excluding overly wide configurations ($k = 7,
10$) that are rarely practical in large-batch regimes, the average
improvement increases to 28\%. A fixed $(k,d)$ policy is
context-agnostic: it applies the same expansion pattern to every node
regardless of the draft model's local confidence or the verification
cost at the current tree size. SMART, by contrast, makes per-node
expansion decisions based on whether the marginal acceptance benefit
exceeds the marginal drafting and verification cost, allowing the tree
to grow deeper on easy contexts and to be pruned earlier on hard ones.
 
\end{document}